\documentclass[11pt]{article}
\usepackage{amsmath,amssymb,graphicx,apacite,color,epsfig,setspace}

\begin{document}

\title{Stimulus-dependent correlations in threshold-crossing spiking neurons\footnote{\textit{Neural Computation},
in press. Received July 30, 2008; accepted February 9, 2009.}}

\date{July 30, 2008}
\author{Yoram Burak$^{1}$, Sam Lewallen$^{2}$, and Haim Sompolinsky$^{1,3}$\\
$^{1}${Center for Brain Science, Harvard University}\\
$^{2}${Faculty of Arts and Sciences, Harvard University}\\
$^{3}${Interdisciplinary Center for Neural Computation, Hebrew University}}
\maketitle
\begin{abstract}We consider a threshold-crossing spiking process as a simple model for the activity within 
a population of neurons. Assuming that these neurons are driven by a common fluctuating input with Gaussian statistics, 
we evaluate the cross-correlation of spike
trains in pairs of model neurons with different thresholds. This correlation function tends to be 
asymmetric in time, indicating a preference for the neuron 
with the lower threshold to fire before the one with the higher threshold, 
even if their inputs are identical. 
The relationship between these results and spike statistics in other models of neural
activity are explored. 
In particular, we compare our model
with an integrate-and-fire model in which the
membrane voltage resets following each spike. 
The qualitative properties of spike cross-correlations,
emerging from the threshold-crossing model, are similar to those of bursting events
in the integrate-and-fire model. This is particularly true for generalized integrate-and-fire models in which spikes
tend to occur in bursts as observed, for example, in retinal ganglion cells driven by 
a rapidly fluctuating visual stimulus.
The threshold
crossing model thus provides a simple, analytically tractable description of event onsets in these neurons.
\end{abstract}
\section{Introduction}

Probing 
the relationship between a stimulus and the spike train, generated by a population
of neurons, is a central theme in the study of neural activity \cite{Spikes}.
Correlations in the spike timing of different neurons are of 
interest, in this context, for two main reasons: First,
spike correlations are informative about the structure of the neural code, 
beyond what could be inferred
from the firing properties of single neurons alone.
Second, correlations are often thought
to reflect the structural properties of the neural network
\cite{Ginzburg1994}. 

Correlated firing in a pair of neurons can arise because
of two distinct reasons:
the existence of a connection
between the neurons (direct or indirect), and the existence of
a common input to the two neurons that is varying in time.
Here we focus on the second possible source for correlated activity.
We consider model neurons 
that receive an analog, continuous stimulus, and respond
to it by generating
a discrete sequence of spiking events. The main question that we address is
how the statistical properties
of the fluctuating stimulus affect the structure of spike correlation functions.

Correlated inputs to neurons were
considered theoretically mainly in context of their effect on single-neuron properties,
such as the firing rate \cite{Tuckwell1988b,SalinasSejnowski2000,MorenoRochaRenartParga2002,KuhnAertsenRotter2003}, 
the coefficient of variation (CV) 
\cite{Tuckwell1988b,BrunelSergi1998,SalinasSejnowski2000,StroeveGielen2001,SalinasSejnowski2002,SchwalgerSchimanskyGeier2008},
and the spike-triggered average stimulus
\cite{KanevWenningObermayer2004,Badel2006,Paninski2006}.
These works considered integrate-and-fire model neurons \cite{Tuckwell1988a,Gerstner}. 

The non-leaky integrate-and-fire model
is relatively tractable analytically \cite{Tuckwell1988b,Paninski2006}. 
In comparison, analytical
treatment of the leaky integrate-and-fire (LIF) neuron is considerably more difficult \cite{Burkitt2006}. Some analytical
results are available for the firing rate
\cite{SalinasSejnowski2000,MorenoRochaRenartParga2002,KuhnAertsenRotter2003},
whereas analytical results for the ISI and the CV are available only in particular limits such as 
slowly varying \cite{MorenoParga2006,SchwalgerSchimanskyGeier2008},
or binary \cite{SalinasSejnowski2002}, inputs. Other results were obtained for the LIF model
from computer simulations \cite{SalinasSejnowski2000,StroeveGielen2001}. 
In particular, \citeA{StroeveGielen2001} included a simulation study of correlations
in spiking of LIF neurons that receive partly overlapping input. Correlations due to partly
overlapping input were also studied analytically,
 but only in the limit where
 the input is slowly fluctuating in time \cite{MorenoParga2006}.

Here we consider a simpler model of neural response, where neurons spike
whenever a generating potential $g(t)$, linearly related
to the neuron's stimulus, crosses a threshold in its rising phase. 
For sufficiently simple
stimuli this model is analytically tractable,
which allows for spike correlation functions to be evaluated in closed form.

Threshold crossing processes without a reset were previously analyzed as models
for neural firing.
The spike auto-correlation function
of a neuron was considered by \citeA{Jung1994}, 
while making specific assumptions on the nature of the fluctuating potential.
Here we evaluate the auto- and cross-correlation in spike timing of neurons with
different thresholds, while making fewer assumptions on the generating potentials
eliciting the spikes. These are assumed to be Gaussian, and may be
identical or partially overlapping in different neurons.

As in the case of Linear-Nonlinear (LN) models 
\cite{Korenberg1986,Chichilnisky2001}, we assume that a neuron responds
to a temporal convolution of its stimulus with a linear kernel.
However, in the LN model spiking is stochastic,
whereas in the threshold-crossing model the spike timing
is precisely determined by the stimulus. This aspect of the model
is motivated by the observation, in various neural assays, 
of responses 
that are precisely repeatable across multiple trials
\cite{MainenSejnowski1995,Berry1997,Bialek1997,MeisterBerry1999,Predicting,UzzellChichilnisky2003}, much more than
can be described by Poisson statistics,
particularly when the stimulus is strongly fluctuating in time.

The model and our main assumptions are presented in Sec.~II. Before discussing 
spike correlation functions, we
first  evaluate
the firing rate (Sec. III), the spike-triggered average stimulus, and the
spike-triggered covariance (Sec. IV). We then evaluate spike auto- and cross-correlation 
functions (Sec.~V). These results are compared with computer simulations of model neurons
that spike according to several variations of the leaky integrate-and-fire model  (Sec. VI).

\section{Formalism and assumptions}

We consider one or more neurons that respond to the same 
generating potential $g$ and elicit a spike whenever $g$ crosses
a threshold in its rising phase. 
The spike train generated by the neuron $i$ can be written
as 
\begin{equation}
\chi_i(t) = \delta\left[g(t)-\theta_i\right] \dot{g}(t) \Theta\left[\dot{g}(t)\right]
\label{eq:chi}
\end{equation}
where $g$ is the generating potential. 
The Heaviside step function $\Theta\left[\dot{g}(t)\right]$ restricts the firing
to the upward crossing events. 

We assume that $g$ is stationary, Gaussian, and has zero mean. 
The 
generating potential is
thus fully characterized by its correlation function,
\begin{equation}
\left<g(t)g(t')\right> = w(|t'-t|)
\end{equation}
where the brackets $\left<\right>$ stand for an ensemble average over all possible 
realizations of the fluctuating generating potential $g$.
Later on, in Sec.~V, we consider a more general situation of a population of neurons, whose generating potentials 
$g_i$ are jointly Gaussian, characterized by their covariance functions
\begin{equation}
\left<g_i(t)g_j(t')\right> = w_{ij}(t'-t).
\end{equation}
The
properties of a model neuron, responding to a stimulus whose mean differs from zero, can be obtained
from the zero-mean case simply by adjusting the threshold.

For a single neuron, the behavior of $w(\Delta t)$ at small $\Delta t$ determines the firing rate and the 
short-time behavior of the spike auto-correlation (Sec.~V). 
We assume that an
expansion of $w(\Delta t)$ exists around $\Delta t = 0$,
\begin{equation}
w(\Delta t) = W_0 + W_1 |\Delta t| + \frac{1}{2} W_2 \Delta t^2 + \frac{1}{3!} W_3 |\Delta t|^3 + \ldots
\label{eq:w_expand1}
\end{equation}
We further assume that $W_1$ vanishes and that $W_2$ is negative, for reasons that will 
become clear later on. As seen below, it is necessary to
treat separately different classes of processes, depending on which other coefficients in the expansion are non-zero.
One important class is the case where $W_3 \neq 0$. 
This is the typical situation if a causal filter is involved in the generation of $g(t)$.
For comparison we also briefly consider, in Sec. V,
the case where all the coefficients with odd indices vanish; 
for example, a process with a Gaussian correlation function, $w(t) = \sigma^2{\rm exp}(-t^2/2 \tau^2)$.

\subsection*{Relationship between $g$ and the stimulus}

The generating potential in our model is linearly related to the stimulus $s$,
\begin{equation}
g = f \circ s 
\label{eq:g_f}
\end{equation}
where the symbol `$\circ$' stands for convolution. We assume that 
$s$ is a zero-mean, uncorrelated Gaussian process:
\begin{equation}
\left<s(t) s('t)\right> = \sigma_0^2 \delta(t'-t)
\end{equation}
The correlation function of $g$ is then  
\begin{equation}
w(\Delta t) = \sigma_0^2  \int f(t) f(t+\Delta t) {\rm d}t
\label{eq:wf}
\end{equation}
Note that $W_0 = \sigma_0^2 \int_{-\infty}^{\infty} \left[f(t)\right]^2 {\rm d}t  > 0$ and that if $f$ is sufficiently regular, 
$W_2 = -\sigma_0^2 \int_{-\infty}^{\infty} \left[f'(t)\right]^2 {\rm d}t < 0$.

The assumption that $s$ is uncorrelated is made 
for simplicity of the presentation, and because uncorrelated flickering stimuli
are often used experimentally. 
The results below generalize in a straightforward manner also to the case where $s$ is a correlated
Gaussian process. 

\subsection*{Specific forms for $f$ and $w$}

In the following sections we first derive results that are
valid generally, for any $w(\Delta t)$. We then illustrate these results with a specific 
example, where we assume a particular form of $f$.
In these examples the filter $f$ is a causal filter of the form
\begin{equation}
f(t) = \left\{ \begin{array}{ll} 0 & \ \ \ \ t < 0 \\
{\displaystyle \frac{{\rm e}^{-t/\tau_2} - {\rm e}^{-t/\tau_1}}{\tau_2-\tau_1}}& \ \ \ \ t > 0 \end{array}\right.
\label{eq:f}
\end{equation}
which is a combination of two single-exponential 
filters with time constants $\tau_1$ and $\tau_2$. Using Eq.~(\ref{eq:wf})
\begin{equation}
w(\Delta t) = \sigma^2 \frac{\tau_2 {\rm exp}\left(-|\Delta t|/\tau_2 \right) - \tau_1 {\rm exp}\left(-|\Delta t|/\tau_1\right)}{\tau_2-\tau_1}
\label{eq:w}
\end{equation}
where 
\begin{equation}
\sigma^2 = \frac{\sigma_0^2}{2(\tau_1+\tau_2)}
\end{equation}
is the variance of $g$.

In the particular case where $\tau_1 = \tau_2 \equiv \tau$, $f$ is an `alpha' filter,
\begin{equation}
f(t) = \left\{ \begin{array}{ll} 0 & \ \ \ \ t < 0 \\
\displaystyle
\frac{t}{\tau^2} {\rm e}^{-t/\tau} & \ \ \ \ t > 0 \end{array}\right.
\label{eq:falpha}
\end{equation}
and
\begin{equation}
w(\Delta t) = \frac{\sigma^2}{\tau} \left(|\Delta t|+\tau\right) {\rm e}^{-|\Delta t|/\tau}
\label{eq:walpha}
\end{equation}
The expansion of $w(\Delta t)$ for small $\Delta t$, 
Eq.~(\ref{eq:w_expand1}), yields $W_1 = 0$ and:
\begin{equation}
W_0 = \sigma^2 \ \ , \ \ 
W_2 = -\frac{\sigma^2}{\tau_1 \tau_2} \ \ , \ \
W_3 = \frac{\tau_1 + \tau_2}{\tau_1^2 \tau_2^2} \sigma^2\ \ , \ \
W_4 = -\frac{\tau_1^2 + \tau_1 \tau_2 + \tau_2^2}{\tau_1^3 \tau_2^3}\sigma^2 \ \ , \cdots
\label{eq:W_double_f}
\end{equation}

\section{Firing rate}

We begin with the relatively 
simple problem of evaluating the firing rate of a single neuron \cite{Rice},
\begin{equation}
r = \left<\chi(t)\right>
\end{equation}
where $\chi$ is given by Eq.~(\ref{eq:chi}). 
Although this quantity has been calculated before, we derive it here in some detail,
because the derivation generalizes to the higher-order moments that are calculated later (whose
detailed derivation is presented in the appendices.) The firing rate $r$ can be written as
\begin{equation}
r = \int_0^{\infty} {\rm d}q\, q\cdot p(\theta, q) 
\label{eq:firing_rate}
\end{equation}
where $p(g, q)$ is the joint probability distribution for the generating potential to be equal to $g$ and for its derivative, at the same time, to be equal to $q$. To evaluate this and similar quantities, we use an identity that holds for a general Gaussian signal $g(t)$
with correlation function $w(\Delta t)$:
If $\zeta_1\ldots \zeta_n$ are all scalar random variables that depend linearly on the process $g$,
\begin{equation}
\zeta_i = \int {\rm d}t \, \alpha_i(t) g(t),
\label{eq:zeta}
\end{equation}
then the joint probability distribution of $\zeta_1,\cdots,\zeta_n$ is equal to
\begin{equation}
p(\zeta_1,\cdots,\zeta_n) = \frac{1}{Z}\,
{\rm exp}\left[-\frac{1}{2}{\bf \zeta}^T A^{-1} {\bf \zeta}\right]
\label{eq:zeta_p}
\end{equation}
where $Z = \left[(2\pi)^n {\rm det}A\right]^{1/2}$ and
\begin{equation}
A_{ij} = \int {\rm d}t \int{\rm d}t'\, \alpha_i(t) w(t'-t) \alpha_j(t')
\label{eq:zeta_A}
\end{equation}

The generating potential and its derivative at time $t = 0$ correspond to $\zeta_1 = g(0)$ and $\zeta_2 = \dot{g}(0)$, therefore
$\alpha_1  =  \delta(t)$ and $\alpha_2  =  -\dot{\delta}(t)$, so that
\begin{equation}
A = \left[\begin{array}{cc} W_0 & 0 \\ 0 & -W_2 \end{array}\right]
\label{eq:A_2}
\end{equation}
The off-diagonal terms are proportional to $w'(0)$, which vanishes if $W_1 = 0$, as we assumed\footnote{If $W_1 \neq 0$, $w(\Delta t)$ does not have a derivative at $\Delta t = 0$, and the derivation performed here fails. A typical example is
given by uncorrelated noise passed through a single-exponential filter. For this process the firing rate diverges (see below).}. We
thus have
\begin{equation}
p(g,q) = \frac{1}{Z}
 {\rm exp} \left(
-\frac{g^2}{2W_0} -\frac{q^2}{-2 W_2}\right)
\label{eq:p_sq}
\end{equation}
and the integral (\ref{eq:firing_rate}) yields
\begin{equation}
r = \frac{1}{2\pi} \left[\frac{-W_2}{W_0}\right]^{1/2}{\rm exp}\left(-\frac{\theta^2}{2 W_0}\right) 
\label{eq:r_eval}
\end{equation}
Hence $W_2$ must be negative for the firing rate to be finite.

For the generating potential of Eqs.~(\ref{eq:f}) and (\ref{eq:w}),
\begin{equation}
r = \frac{1}{2\pi \left(\tau_1 \tau_2\right)^{1/2}} {\rm exp}\left(-\frac{\theta^2}{2 \sigma^2}\right).
\label{eq:r_specific}
\end{equation}
If either $\tau_1$ or $\tau_2$ vanish, the firing rate diverges. Hence
a generating potential obtained by convolving uncorrelated noise with a single exponential filter 
has an infinite rate of threshold crossings. In such a process there
are finite intervals in which the number of crossings is infinite, and others
in which no firing occurs\footnote{This is related to the short-time properties of the spike
auto-correlation function (Sec.~V) in the limit where $\tau_1$ or $\tau_2 \rightarrow 0$.}.

\section{Spike triggered average stimulus}
The spike triggered averaged stimulus (STA) 
is the mean stimulus given that a spike was generated at a particular time. Because the input is assumed to be stationary, the STA is only a function of the time difference relative to the spike time, 
\begin{equation}
f_{\rm STA}(\Delta t)  =  \frac{1}{r}\left< s(\Delta t) \cdot \chi(0)\right> 
\label{eq:STA_def}
\end{equation}
where $r$ is the firing rate, $\chi$ is determined by $g$, as in Eq.~(\ref{eq:chi}), and the spike time was arbitrarily chosen to
be zero.
If one assumes that spiking is described by a LN model with filter $f$, and if the stimulus is uncorrelated Gaussian noise, 
$f_{\rm STA}(\Delta t)$ is proportional to $f(-\Delta t)$ \cite{Chichilnisky2001, Paninski2003}.
Hence it is common to probe the spatio-temporal filter applied by a neuron on its stimulus by measuring 
$f_{\rm STA}$ \cite{Schwartz2006}, 
in particular in the visual sensory system \cite{Rieke2001,Chichilnisky2002,Baccus2002,Rust2005,Hosoya2005}.

To compute $f_{\rm STA}$ for a threshold-crossing spiking neuron
we need the joint probability distribution function $p(g_1, q_1; s_2)$
for $g(0) = g_1$, $\dot{g}(0) = q_1$, and $s(\Delta t) = s_2$, 
evaluated in Appendix A. The STA is then given by
\begin{equation}
f_{\rm STA}(\Delta t) =   \frac{\theta}{A} f (-\Delta t) + \sigma_0 \sqrt{\frac{\pi}{2B}}
f' (-\Delta t)
\label{eq:sta_eval}
\end{equation}
where $A = W_0  /  \sigma_0^2 =  \int_{-\infty}^{\infty} f^2(t) {\rm d}t$ and 
$B = -W_2  /  \sigma_0^2 = \int_{-\infty}^{\infty} f'^2(t) {\rm d}t$. 

The first term in Eq.~(\ref{eq:sta_eval}) is proportional to the expectation for the STA in a rate-based linear-nonlinear model \cite{Chichilnisky2001}. However, for a threshold-crossing spiking neuron it correctly describes the STA only in the limit of large $\theta$: 
in contrast, when $\theta = 0$, the STA is proportional to $f'(\Delta t)$.  

The result that $f_{\rm STA}$ is a linear combination of the filter and its derivative has a simple geometric interpretation, which is most easily seen by thinking of the stimulus history as an $n$-dimensional discrete vector. The condition for spiking at time $t = 0$ involves $g$ and its derivative, which are the inner products of the stimulus history with two vectors, $f$ and $f'$. Because $s$ is
uncorrelated, its history vector is distributed in a radially symmetric manner in $n$-dimensional space.  The STA must lie in the two-dimensional sub-space spanned by the vectors $f$ and $f'$.

This result is illustrated in Fig.~1a, for the generating potential of Eqs.~(\ref{eq:falpha}) and (\ref{eq:walpha}).
With the threshold varying from 0 to $2.5 \sigma$, the shape of $f_{\rm STA}$ varies significantly. The dotted lines in the figure (overlapping with the solid line) were obtained from a discrete simulation, using a time step ${\rm d}t = \tau/100$ and averaging over a simulation run
$T/\tau = 10^6$ for $\theta = 0$ and $\theta = 1$, and $T/\tau= 10^8$ for $\theta = 2.5$.

It is interesting to compare these results with what is expected from a leaky integrate-and-fire model neuron
\cite{Tuckwell1988a,Gerstner}. For this case there is no known analytic form for the STA (see, however,~\citeNP{KanevWenningObermayer2004,Badel2006,Paninski2006}), but we can evaluate the STA numerically. We assume that the membrane potential of the model neuron is related to the input current via 
\begin{equation}
\tau_1 \frac{{\rm d}u}{{\rm d}t} = -u + I
\label{eq:IAF1}
\end{equation}
with a reset of the membrane potential to $u_r$ whenever $u$ reaches the threshold $\theta$, whereas the
input current is related to the stimulus $s$ by
\begin{equation}
\tau_2 \frac{{\rm d}I}{{\rm d}t} = -I + s
\label{eq:IAF2}
\end{equation}
The sub-threshold dynamics of $u$ depends on the stimulus in the same way as the generating potential $g$ depends on stimulus
in the threshold-crossing
model. The difference between the two models lies in the existence of a reset. For simplicity, 
we set $u_r = 0$ and $\tau_1 = \tau_2$ for the rest of this section
(see also Sec.~\ref{sec:iaf} and Appendix D).
 
Figure 1b shows a comparison between $f_{\rm STA}$ in the integrate-and-fire model (simulation, dotted line: 
$dt/\tau = 0.01$ and $T/\tau = 10^6$)
and $f_{\rm STA}$ in the threshold-crossing spiking model [Eq.~(\ref{eq:sta_eval}), solid line.] 
The parameters are as in panel a, with a threshold $\theta = 2.5 \sigma$.  The two models agree very well in their
prediction and differ, significantly, from that of a LN model, where $f_{\rm STA}(\Delta t) \propto f(-\Delta t)$ 
(dashed line).

For a lower threshold, $\theta = 0.25$, the threshold-crossing spiking model and the integrate-and-fire model 
no longer yield similar predictions for $f_{\rm STA}$ (solid and dashed lines, Fig.~1c). The difference
between the cases of high and low threshold is possibly related  to the typical inter-spike interval
which, in the case of a high threshold, is longer, allowing the input to decorrelate between
subsequent spikes. The statistics of the input  before a spike are
thus less influenced, in the case of high threshold,
by whether a reset has occurred following the previous spike.

Finally, we note that the threshold-crossing model is similar to a two-dimensional LN
model involving two filters, equal to $f(-t)$ and $f'(-t)$. In the analogous LN model
the nonlinear transfer function must be chosen 
to be narrowly peaked at values of the generating potential close to the threshold, and an appropriate
dependence on the derivative must be included as well\footnote{There is a difference
between the two models, however, in that the two-dimensional LN model produces a variable number of spikes within the narrow
spiking event.}.
The result that the STA is a linear combination of the filter $f$ and its derivative thus agrees with the
general property of multi-dimensional LN models, that the STA lies within the subspace
spanned by the linear kernels determining the spiking rate 
\cite{Paninski2003,Schwartz2006}. Similarly, we may expect the spike-triggered covariance \cite{BialekVanSteveninck1988,Schwartz2006}
to reveal the linear filter $f$ and its derivative $f'$ as spanning this sub-space.
This is indeed the case (Appendix A.)

\section{Spike correlations}

We next consider correlations in the spike timing of two neurons, assuming that they receive identical inputs,
but possibly differ in their thresholds $\theta_{1,2}$.
Because the generating potential is assumed to be stationary, the spike correlation function depends only on the time difference between spikes:
\begin{equation}
c(\Delta t) = \left< \chi_1(0) \cdot \chi_2(\Delta t)\right >  
\label{eq:sqpike_corr_def}
\end{equation}
From the definition of $\chi_{1,2}$, Eq.~(\ref{eq:chi}), the quantity inside the brackets depends on the 
generating potential and its derivative at $t = 0$ and at 
$t = \Delta t$.

The detailed form  of $c(\Delta t)$ is derived in appendix B, and here
we present the main results.
The spike correlation function can be written as
\begin{eqnarray}
c(\Delta t)   & =  & p(\theta_1;\theta_2) \cdot
 \int_0^\infty {\rm d}q_1 \int_0^\infty {\rm d}q_2\,q_1 \cdot q_2 
\label{eq:I_def} \\
& \times &
\frac{\left({\rm det}\,m^{-1}\right)^{1/2}}{2\pi} 
{\rm exp}\left[-\frac{1}{2}\left({\bf q} - {\bf q}_0\right)^Tm^{-1}\left({\bf q}-{\bf q}_0\right)\right]
\nonumber
\end{eqnarray}
where $p(\theta_1; \theta_2)$ is the joint probability distribution for $g(0) = \theta_1$ and $g(\Delta t) = \theta_2$
[see Eq.~(\ref{eq:p_tt})], and ${\bf q}^T = (q_1, q_2)$.
The quantity in the second line of the equation is the conditional probability distribution function
for $g'(0) = q_1$ and $g'(\Delta t) = q_2$, given that $g(0) = \theta_1$ and $g(\Delta t) = \theta_2$. This
conditional distribution is Gaussian and is characterized by its mean ${\bf q}_0$ and the covariance matrix
$m$, whose values are derived in the appendix [Eqs.~(\ref{eq:m}) and (\ref{eq:q0})].

\subsection*{The spike auto-correlation function}

We briefly discuss the spike auto-correlation function, corresponding to the case $\theta_1 = \theta_2 \equiv \theta$,
and focusing on the behavior at small $\Delta t$. Because the spike train
$\chi(t)$ is a point process, its auto-correlation function necessarily includes a contribution 
\begin{equation}
c(\Delta t) = r \delta(\Delta t) + \ldots
\label{eq:delta_contribution}
\end{equation}
The discussion here concerns nonzero values of $\Delta t$, \textit{i.e.}, the occurrence of spikes in addition to the one at $t = 0$.  The behavior for small $\Delta t$ is remarkably different for the two classes of generating-potential statistics discussed in Sec.~II: If $W_3 \neq 0$, as in the example of Eq.~(\ref{eq:W_double_f}),
the spike auto-correlation function tends to a finite value when $\Delta t \rightarrow 0$:
\begin{equation}
c(\Delta t) \rightarrow \frac{W_3}{4\pi^2 (-3W_0 W_2)^{1/2}}
{\rm exp}\left(-\frac{\theta^2}{2W_0}\right)
\label{eq:c_deltat_auto_w3}
\end{equation}
It is interesting to look at this quantity divided by $r$, which represents the firing rate at time $t = \Delta t$ after the occurrence of a spike at $t = 0$:
\begin{equation}
\frac{c(\Delta t)}{r} \rightarrow \frac{W_3}{-2\pi\sqrt{3} W_2}
\end{equation}
This ratio does not depend on $\theta$. By comparing with $r$ itself we see that, if the threshold is high,  the spike-conditioned firing rate is much higher than the average firing rate. In other words, once the neuron has fired, it is likely to soon fire again, compared to its baseline firing rate. When the threshold is zero, the spike-conditioned firing rate can be larger or smaller than $r$, depending on the expansion coefficients $W_2$ and $W_3$. This result is illustrated in Fig.~2a, where $w$ is given by Eq.~(\ref{eq:walpha})\footnote{
Note that the conditional firing rate diverges if $\tau_1 \rightarrow 0$ or
if $\tau_2 \rightarrow 0$, which can be seen from Eq.~(\ref{eq:W_double_f}). 
This is consistent with the divergence of the average firing rate in this limit, and the existence of 
finite time intervals in which the
number of spikes is infinite.}.

We note that when $W_3 \neq 0$ as above, the second derivative of the generating potential
is an unbounded random process, and
the first derivative of $g$ is not smooth. 
This facilitates the generation of successive spikes at short intervals,
since the derivative of $g$ is required to be positive at the interval's edges, and negative 
somewhere in between.

In contrast, when the first derivative
of $g$ is a smooth process, 
we may expect the occurrence of spikes at vanishingly small intervals to be
considerably less likely. Indeed, if $w(\Delta t)$ has no irregularities at $\Delta t = 0$,
so that all $W_i$ with odd indices vanish,
the occurrence of spikes separated by short intervals is strongly suppressed,
 $c(\Delta t) \sim (\Delta t)^4$ [Appendix C, Eqs.~(\ref{eq:p_tt_same}) 
and (\ref{eq:I_tt_same_approx})]. 
This is illustrated in Fig.~2b for a generating potential with a Gaussian correlation function, $w(\Delta t) = \sigma^2{\rm exp}(-\Delta t^2/2 \tau^2)$. Note that $W_0$ and $W_2$ are the same in the two parts of Fig.~2; consequently, the mean firing rates are identical in these two cases. 

\subsection*{Spike cross-correlation}

We next consider the spike cross-correlation function of two neurons with different thresholds, firing in response to 
the same stimulus.
We note, first, that unless $\theta_1 = \theta_2$ the correlation function is not symmetric with respect to replacement of $\Delta t$ by $-\Delta t$ or, equivalently, with respect to exchange of the two neurons ($\theta_1\leftrightarrow \theta_2$). The generating potential is required to have a positive derivative at $t = 0$ and at $t = \Delta t$, and therefore we may expect a higher probability for joint spiking if the higher of the two thresholds is set at the later time (see Fig. 3). This is indeed the case -- as demonstrated in Fig.~4a
for the generating potential described by Eq.~(\ref{eq:walpha}), and with thresholds set as $\theta_1 = 0.8 \sigma$ and $\theta_2 = \sigma$.
The preference for one neuron to fire after the other neuron is particularly prominent at short time scales of order $\tau$, but it also has a signature at large time scales (Appendix B). 

It is instructive to compare the threshold-crossing model with the prediction of a one-dimensional
linear-nonlinear (LN) model
\cite{Chichilnisky2001}. 
For this comparison, we assume that the linear filter of the LN model is the same as the filter $f$ in 
the threshold-crossing model, and that in both models the stimulus is uncorrelated Gaussian noise.
In the LN model the spike correlation function depends also on the choice of the non-linear 
function applied
to the outcome of the linear filter.
Two particular choices are made in the examples shown in Fig.~4b: The first is linear rectification, $\phi(x) \propto \Theta(x-\theta)$ 
where $\Theta(x) = x$ for $x > 0$ and $\Theta(x) = 0$ for $x < 0$, with thresholds $\theta_1, \theta_2$ chosen
as in Fig.~1a (solid line). 

The second example (dashed line) is the case where $\phi(x)$ is non-zero only in
a narrow range around $\theta$,
$\phi(x) \propto \delta(x-\theta)$. 
Mathematically, this example compares more directly with the threshold-crossing model, because
the LN model produces spikes only at the threshold crossings.
The correspondence can be seen in the spike cross-correlation function $c(\Delta t)$
obtained from the LN model in this case, which is  
calculated in Appendix E [Eq.~(\ref{eq:c_LN_general}).] In the limit of large $\Delta t$,
 $c(\Delta t)/r_1 r_2 - 1$ in the LN model
[Eq.~(\ref{eq:c_LN_delta})] 
is equal to the first term in the large $\Delta t$ expansion for
the threshold-crossing model,
Eq.~(\ref{eq:spike_corr_large_t}). 

We note two important qualitative differences between the LN model and the
threshold-crossing spiking model: First, in the LN model, the spike correlation function is symmetric under time reversal:  
$c(-\Delta t) = c(\Delta t)$\footnote{This property arises for the following reason. The spike cross-correlation at any two times $t_1, t_2$
depends on the joint probability distribution of the generating potential values at these times. Assuming
that the generating potential is Gaussian,
this joint distribution function is necessarily symmetric.}. 
A second qualitative difference between the two models, seen in Fig.~4, is that the firing of the two neurons is less correlated in the LN model,
compared to the threshold crossing spiking model, at short time scales.

\subsection*{The limit $\Delta t \rightarrow 0$}

In the limit of small $\Delta t$ the behavior of $c(\Delta t)$ is substantially different if the
neuron with the higher threshold fires before or after the neuron with the lower threshold\footnote{A 
weak asymmetry between positive and negative $\Delta t$ exists also in the
limit $\Delta t \rightarrow \infty$. This limit is considered in Appendix B.}. To 
simplify notation we assume here that $\theta_2 > \theta_1$.

$\mathbf{\Delta t > 0}$: For small $\Delta t$
the spike correlation function scales with $\Delta t$ as
\begin{equation}
c(\Delta t) \sim \frac{(\theta_2-\theta_1)^2}{\Delta t^3}
{\rm exp}\left[\frac{1}{2W_2}\left(\frac{\theta_2-\theta_1}{\Delta t}
\right)^2
+ \frac{\gamma_1}{\Delta t}
\right]
\label{eq:p_tt_small_t_approx}
\end{equation}
where
\begin{equation}
\gamma_1  =  -\frac{W_3}{6 W_2^2}(\theta_2-\theta_1)^2
\end{equation}
This approximation, with the appropriate prefactors taken from Eqs.~(\ref{eq:c_p_I}), (\ref{eq:I_small_t_pos_ap}),
and (\ref{eq:p_tt_small_t}) is plotted in Fig.~4a (dashed line) for the particular
input considered in this figure.

As $\Delta t \rightarrow 0$, the exponential decay in Eq.~(\ref{eq:p_tt_small_t_approx}) 
wins over the $\sim(\Delta t)^{-3}$ divergence for sufficiently small $\Delta t$. At larger values of $\Delta t$ 
the spike correlation function peaks, and then decays algebraically. Hence the neuron with the larger threshold tends
to fire after the neuron with the lower threshold, with a typical latency given by the position of the peak.
If we ignore the term 
$\gamma_1/\Delta t$, which is legitimate when
the difference in thresholds is sufficiently small,
the maximum is at $\Delta t = \Delta t^*$ where
\begin{equation}
\Delta t^* = \frac{\theta_2-\theta_1}{\sqrt{-3W_2}}
\label{eq:latency}
\end{equation}
For the generating potential of Eqs.~(\ref{eq:f}) and (\ref{eq:w}),
\begin{equation}
\Delta t^* = \frac{\theta_2-\theta_1}{\sigma} \sqrt{\frac{\tau_1 \tau_2}{3}}
\label{eq:latency_tau}
\end{equation}

Equation (\ref{eq:latency}) indicates that the most likely latency increases linearly with $\theta_2 - \theta_1$. 
An illustration of this result is shown in Fig.~5, where $f$ and $w$ are as defined by Eqs.
(\ref{eq:falpha}) and (\ref{eq:walpha}), 
$\theta_2 = 0.5 \sigma$, and $c(\Delta t)$ is plotted 
for three values of $\theta_1$:  $0.2 \sigma$, $-0.2 \sigma$, and $-0.5 \sigma$ (solid lines). With increase in $\theta_2 - \theta_1$ the peak of the spike correlation function becomes wider, and occurs at larger latencies. The arrows indicate the prediction for the position of the peak, Eq.~(\ref{eq:latency_tau}), which matches the actual position very well even when $\theta_2 - \theta_1$ is relatively large. 

$\mathbf{\Delta t < 0}$:  In this case the process $g$ is required to have a positive derivative at $t = \Delta t < 0$, a negative derivative within the interval $(\Delta t, 0)$ and, again, a positive derivative at $t = 0$. 
Such a trajectory becomes increasingly unlikely when $\Delta t \rightarrow 0$, and the spike correlation function decays to zero. The leading contribution is of the form
\begin{equation}
{\rm log}c(\Delta t) \simeq -\frac{3(\theta_1-\theta_2)^2}{W_3(\Delta t)^3} + \cdots
\label{eq:log_c_deltat}
\end{equation}
if $W_3 \neq 0$. If $W_3 = W_5 = 0$ the decay is even stronger, ${\rm log} c(\Delta t) \propto (\Delta t)^{-6}$ (Appendix C.)
Equation~(\ref{eq:log_c_deltat}) introduces a characteristic time scale for the inhibition, scaling
as 
\begin{equation}
\frac{\left(\theta_1-\theta_2\right)^{2/3}}{W_3^{1/3}}  = \left(\frac{\tau_1^2 \tau_2^2}{\tau_1+\tau_2}\right)^{1/3}
\left(\frac{\theta_1-\theta_2}{\sigma}\right)^{2/3}
\end{equation}
where the expression on the right hand side holds if $w(\Delta t)$ is given by Eq.~(\ref{eq:w}).

\subsection*{Partly overlapping inputs}

We next generalize the
analysis to the more realistic case where the inputs to different neurons are only partially overlapping:
\begin{equation}
g_i = f \circ (s + \xi_i)
\label{eq:g_f_s_xi}
\end{equation}
The noise inputs $\xi_i$ are assumed to be zero-mean, jointly Gaussian processes that are uncorrelated with $s$ and with each other:
\begin{equation}
\left<\xi_i(t) \xi_j(t')\right>  = \alpha \sigma_0 \delta_{ij} \delta(t'-t).
\end{equation}
Here $\alpha$ is the ratio between the standard deviation of the noise and that of the common stimulus.
The covariance of the generating potentials is then
\begin{equation}
w_{ij}(\Delta t) = w(\Delta t)(1 + \alpha^2 \delta_{ij})
\end{equation}
The spike auto- and cross-correlation functions can be evaluated in a similar manner as
for the noise-free case (Appendix B).

As an example, we consider the `alpha' filter, Eq.~(\ref{eq:falpha}). In 
Fig.~6 the standard deviation $\sigma$ of  the common stimulus   is kept fixed, while
$\alpha$ is varied.
Figure~6a shows the spike correlation function of two model neurons with the same
threshold $\theta = \sigma$ and with weak independent noise, $\alpha = 0.1$.
The existence of noise 
increases the correlation in neural firing at non-zero latencies, compared to the noise-free
case (dashed line). 
To understand this seemingly counter-intuitive result, note that 
the spike correlation function in the noise-free case is identical to the spike auto-correlation
function of a single neuron, and includes a delta-function contribution at
$\Delta t = 0$.
When the two neurons receive independent noise, they
no longer fire precisely a the same time, leading to a broadening of the delta function
into a peak of finite width.

Panel b of the figure shows the spike correlation function of two neurons with different thresholds,
$\theta_1 = 0$ and $\theta_2 = \sigma$, for several values of $\alpha$: 0 (no noise),
0.4, and 1. Increasing $\alpha$ reduces the correlation between the two neurons. Note, however, 
that the preference
of neuron $2$ to fire after neuron 1 is clearly evident
at time scales comparable to $\tau$,
even when the noise and signal have the same standard deviation. 

The inset shows $c(\Delta t)/r_1 r_2 - 1$ in a case where the noise is much
larger than the signal, $\alpha = 10$.
Here $c(\Delta t)$ is very well approximated using a linearization with respect to the cross-correlation 
function (dotted line), as described in Appendix B, Eq.~(\ref{eq:weak_correlation}). This approximation 
works quite well even if $\alpha$ is of order unity,  as can be seen in the main plot 
(Fig.~6b, dotted line.)

\section{Comparison with integrate-and-fire neurons}
\label{sec:iaf}

Having characterized the spike correlation statistics of the threshold crossing model, we may
ask whether the results carry  over to other models of neural firing.  
In even the most simple leaky integrate-and-fire (LIF) 
model an analytical form is not known for the spike correlation functions. 
Hence the discussion in this section is based on numerical simulation of model neurons, using
several variations of the LIF model.
We focus on the relative timing
of spikes elicited by neurons that receive the same stimulus but differ in their thresholds.

In the following examples,
where we consider pairs of neurons that differ in $\theta$, we suppose that the neurons
are identical, but differ in the mean value of their fluctuating 
input currents. Hence, after shifting the membrane potential 
to compensate for the different baseline currents, the reset potentials $u_r$ 
in the two neurons differ,
 but the difference $\theta-u_r$ is fixed.

Figure 7\,a shows the spike correlation function of two such neurons, evaluated from
simulation of model neurons according to the LIF model of Eqs.~(\ref{eq:IAF1})--(\ref{eq:IAF2}),
with $\theta_1 = 0.8\sigma$, $\theta_2 = \sigma$, and $\theta - u_r = 0.5 \sigma$. 
The only difference between the LIF model and the threshold-crossing model
is the existence of a reset in 
the membrane potential following each spike. 
Nevertheless, comparing Fig.~7\,a with
Fig.~ 4\,a reveals that there is very
little resemblance between the spike correlation functions obtained from
the two models. Most notably,  in the LIF model there is almost no preference for
one of the neurons to fire later than the other one. 
 
To better understand this discrepancy, Fig.~7\,b shows an example of 
spike trains generated by the two neurons in the LIF model (top two traces),
and in the threshold crossing model (bottom two traces).
Compared to the LIF model, which tends to produce bursts of spikes, 
the threshold crossing model 
tends to generate much more isolated spikes.

The inset in Fig.~7\,b shows an example of bursts of spikes, generated by the LIF 
neurons after a relatively long silent period (top two traces). 
The first spike generated by the neuron 
with the smaller threshold (black) precedes the first spike generated by the neuron
with the higher threshold (gray), preserving the tendency that is observed in
the threshold crossing model. This tendency is typically observed also
in other isolated bursts.

Within the bursts, however, there is no clear relative timing of spikes of the two LIF neurons,
because a spike in one neuron can be paired with a spike in the other neuron that
either precedes it or comes after it. This suggests that the existence
of bursts masks
the tendency of the neurons to fire in a particular order.
Motivated by these observations, we consider several situations in which 
the LIF model can exhibit a characteristic order of spike timing, in similarity to
the prediction of the threshold-crossing
model.

\subsection*{Sparse firing}

In situations where the LIF neurons fire isolated spikes, we may expect a clear
order of spike timing to emerge, and to be reflected in the spike correlation function.

\textit{Sparse firing due to a large potential reset, compared to the input variance.} With increase
of $\theta - u_r$, compared to $\sigma$, spikes tend to become more isolated. 
With fixed 
biophysical parameters of the cell, an increase in $(\theta-u_r)/\sigma$ corresponds to 
a decrease in the standard deviation of the fluctuating input current (see, also,
Appendix D). Figure 8\,a shows that, as $(\theta - u_r)/\sigma$ increases
in both neurons, a pronounced asymmetry
develops in their spike cross-correlation function.

\textit{Sparse firing due to refractoriness.}
Sparse firing may result, alternatively, from the existence of an additional refractory 
mechanism. 
There are many possible ways to model 
refractoriness in a LIF model \cite{Gerstner}, and here we consider one such possibility.
After each spike, the membrane potential $u$
is kept fixed at $u_r$ during a waiting period whose length varies randomly
from spike to spike. 
After the waiting period the membrane potential continues
to evolve according to Eq.~(\ref{eq:IAF1}), and
the distribution function of waiting times decays exponentially 
with a time constant $\tau_r$ . Spike correlation functions, obtained from simulations of
this model with $\tau_r = 2 \tau$, are shown in Fig.~8\,b for several pairs of thresholds,
exhibiting a clear asymmetry in firing order. Note that in this example, the membrane time constant
is shorter than the typical refractory time $\tau_r$.

\subsection*{Bursting events}

Neurons often generate distinct bursts of spikes, sometimes referred to as 
events,  when presented with stimuli possessing
rapid temporal fluctuations \cite{Berry1997,MeisterBerry1998,Berry1999}. 
It has been argued that the timing of events, as marked by their first spike,
can convey significant information about the stimulus
\cite{Hopfield1995,GollischMeister2008}.
Hence
it is of interest to understand the factors influencing the relative timing of
these spikes. In the following, we consider a model
where bursting events are more clearly separated from each other than in 
the simple LIF model. We then interpret the predictions of the threshold-crossing
model in relation to the timing of burst onsets.

To reproduce well-separated spiking bursts in a neural model, we consider 
a generalization of the LIF model, introduced by \citeA{Predicting}. This model
was shown to reliably predict the structure of spike trains generated by retinal ganglion cells from several
different species, in response to rapidly flickering stimuli. In a version of this model
without noise, a generating potential $g(t)$ is related to the stimulus by Eq.~(\ref{eq:g_f}), as in the threshold 
crossing model. Spiking occurs when $g(t)$ crosses a time-dependent threshold $b(t)$ that increases
following each spike,
\begin{equation}
b(t) = \theta + B \int_{-\infty}^{t} \chi(t') {\rm exp}\left(-\frac{t-t'}{\tau_p}\right) {\rm d}t'
\end{equation}
We refer to this model as the variable-threshold (VT) model.

We may interpret $g-b+\theta$ as the neuron's membrane potential: with this interpretation
the discrete increase in $b$ following a spike 
corresponds to a reset of the membrane potential from $\theta$ to $\theta - B$. 
The VT model is precisely equivalent to the LIF model of equations (\ref{eq:IAF1})--(\ref{eq:IAF2}) 
if $f$ is a double
exponential filter, as in Eq.~(\ref{eq:f}), and if $\tau_p = \tau_1$. 
For simplicity we take $f$ to be an `alpha' filter, Eq.~(\ref{eq:falpha}). However, we
choose the time scale of threshold recovery $\tau_p$ to be larger than the neuron's membrane time constant,
$\tau_p = 5\tau$. This ratio roughly matches the relation
between $\tau_p$ and the shape of filters that were found in
\citeA{Predicting} to provide a good description of spike trains
from retinal ganglion cells.

A spike train generated by this model is shown in Fig.~9b, and consists of 
clearly separated events. 
The red lines in Fig.~9\,b represent spikes
generated by a threshold-crossing model with the same generating potential 
and threshold as in the VT model. These spikes roughly match the onset of busting events
in the VT model.
The threshold-crossing model thus provides a coarse-grained description of events
generated by the VT model. 

We next isolate the first spikes of events in the VT model by
discarding any spike that occurred within a time delay $D = 2\tau$ from a 
previous spike\footnote{Because the events are clearly separated clearly from each other, the results
are insensitive to the precise choice of $D$.}.
Fig~9\,c shows the cross-correlation function of these
specifically selected spikes in several neurons with thresholds
$\theta/\sigma = 0.5$, $0.2$, $-0.2$, and $-0.5$, that were all presented with the same stimulus. These 
cross-correlation functions display a strong asymmetry between positive and negative
$\Delta t$, and are qualitatively very similar to the spike cross correlations in the threshold crossing model,
shown in Fig.~5. 
In contrast 
 the spike cross correlation functions calculated directly from all 
 the spikes, Fig.~9\,a, are very different from the prediction of the threshold-crossing model.
 
 These results suggest that the onsets of bursting events in neurons that
 receive the same stimulus, but differ in their threshold,  can
 exhibit strong asymmetry in their relative timing, and that the correlation functions of these event
 onsets can be qualitatively described by the simplified threshold-crossing model.
 
\section{Summary}

We considered in this work a relatively simple
deterministic process that produces a discrete spike train from an analog, 
continuous signal. The timing of spikes in this model is more precisely controlled by the stimulus
than typically predicted by rate-based models. This aspect of the model is motivated
by observations of precisely timed spiking in neural assays, particularly in response
to stimuli that
possess strong temporal modulations
\cite{MainenSejnowski1995,Berry1997,Bialek1997,MeisterBerry1999,UzzellChichilnisky2003}.
Some of the salient properties of spike correlation functions in this model are summarized below.

Most notably,
two neurons receiving identical inputs can show a preference for one neuron to fire later than the 
other, although there is no monosynaptic connectivity between them (Figs.~4a and 5.)
This preference is prominent even if the thresholds of the two neurons are similar
(but not equal), and if there is only partial overlap in their input (Fig.~6.)

Comparison with the leaky integrate-and-fire model shows that bursting
in LIF neurons often masks the preference of neurons to fire in a particular order (Fig. 7). However, there are
several situations in which such a preference may be observed in LIF neurons.
First, LIF neurons can display a preference to fire in a particular order
if they produce sparse spiking,
\textit{e.g.}, due to a large hyperpolarizing step in the membrane potential after each spike, 
compared to the standard deviation of the stimulus, 
or due to other sources of refractoriness (Fig. 8).
Second, when the neurons generate clearly separated bursting
events, the first spikes of these events have cross-correlation functions that are
 similar to those
predicted by the threshold crossing model (Fig. 9). 

The spike cross-correlation function of neurons that differ only in their thresholds may be
 probed experimentally by repeatedly presenting the same stimulus to a single neuron,
 while injecting varying amounts of current into the neuron from trial to trial.
 Correlations between different trials can then effectively measure
the spike correlation function of two identical neurons with different thresholds.
 This approach was recently taken in \cite{Markowitz2008}, where a Gaussian stimulus,
 mimicking the spectral properties of Gamma oscillations, was injected into
 rat pyramidal neurons from the somatosensory cortex. While 
 spike cross-correlation functions are not shown in \cite{Markowitz2008}, spike trains 
 from different trials exhibit a clear modulation of spike timing by the varying
 injected current, suggesting that a strong asymmetry exists in the spike-correlation function of
 neurons with different thresholds.
 Furthermore, the characteristic latency between the firing of two neurons
 appears to increase linearly as a function of the difference in injected currents.
 This roughly linear dependence persists
 over a wide range of current differences, in similarity to our results from the
 threshold-crossing model.
 
We also considered in this work the spike-triggered average stimulus (STA).
The STA in our model is a linear
combination of the filter and its derivative. A simple geometrical argument
shows that this result extends to a larger
class of models: it should hold whenever the stimulus is passed through a  
linear kernel, if spike decisions are then based strictly
on the output of the kernel and its derivative.
Varying the threshold
modifies the relative weight of the filter and its derivative (Fig 1).
 It will 
be interesting to probe for such a dependence of the STA on threshold
in real neurons.

\section*{Acknowledgements}
We acknowledge helpful discussion with Robert G\"{u}tig  and with Markus Meister, and
thank Uri Rokni for useful comments on the manuscript.
We thank Markus Meister for sharing with us his unpublished notes
on threshold crossing processes. 
YB acknowledges support from the Swartz foundation. HS is partially supported
by a grant from the ISF and from the Israeli ministry of defense (MAFAT). 

\section*{Note added in proof}
Tchumatchenko et al. have recently considered a threshold-crossing model, similar to the one presented
in this work. A preprint of their work has been made available on the arXiv.org e-Print archive 
while our manuscript was in review \cite{Tchumatchenko2008}.

\newpage

\appendix
  
\section{Spike triggered average stimulus and covariance}

To avoid infinities in the calculation we assume first that $s$ is a
correlated Gaussian process with
a correlation function $w_s(\Delta t)$. At the end of the calculation
we take the limit
\begin{equation}
w_s(\Delta t)  \rightarrow \sigma_0^2 \delta(t-t') 
\label{eq:sigma_s_xi_lim}
\end{equation}
The correlation function of $g$ is 
\begin{equation}
w(\Delta t) = \int {\rm d}t \int{\rm d t'} \,f(t) w_s(t-t'+\Delta t) f(t')
\label{eq:wf_split}
\end{equation}

To evaluate $f_{\rm STA}$ we need the joint
probability distribution function for $g(0) = \theta$,  $\dot{g}(0) = q_1$, and $s(\Delta t) = s_2$, in terms of which
$f_{\rm STA}$ is given by
\begin{equation}
f_{\rm STA}(\Delta t) = \frac{1}{r} \int_{-\infty}^{\infty} {\rm d}s_2 \int_0^\infty {\rm d} q_1 \, s_2\cdot q_1 \cdot p(\theta, q_1; s_2) 
\label{eq:sta_integral}
\end{equation}
The joint probability distribution of $p(\theta,q_1; s_2)$ is given by
\begin{equation}
p(\theta, q_1; s_2) = \frac{1}{Z}\,
{\rm exp}\left[-\frac{1}{2}{\bf \zeta}^T A^{-1} {\bf \zeta}\right]
\label{eq:zeta_p_N}
\end{equation}
where $\zeta^T = (\theta, q_1, s_2)$, $Z = \left[(2\pi)^n {\rm det}A\right]^{1/2}$, and
\begin{equation}
A = \left[
\begin{array}{ccc}
w(0) & 0 & (f\circ w_s)(-\Delta t) \\
0  & -w''(0) & (f' \circ w_s)(-\Delta t)  \\
(f\circ w_s)(-\Delta t) & (f'\circ w_s)(-\Delta t)  & w_s(0) 
\end{array}
\right]
\label{eq:A3}
\end{equation}
Evaluating the integral (\ref{eq:sta_integral}) yields
\begin{equation}
f_{\rm STA}(\Delta t) = \frac{\theta}{w(0)} (f \circ w_s)(-\Delta t) + \sqrt{\frac{\pi}{-2w''(0)}}
(f' \circ w_s)(-\Delta t)
\label{eq:sta_eval2}
\end{equation}
In the limit where $w_s$ is uncorrelated, Eq.~(\ref{eq:sigma_s_xi_lim}), we get 
 Eq.~(\ref{eq:sta_eval}).
 
The spike-triggered covariance $C(t_1,t_2)$ is defined as
\begin{equation}
C(t_1,t_2) = \frac{1}{r} \left<\left [s(t_1) - f_{\rm STA}(t_1)\right]
\left [s(t_2) - f_{\rm STA}(t_2)\right] \cdot \chi(0) \right>
\end{equation}
A calculation similar to that outlined for the STA yields 
\begin{equation}
\frac{C(t_1,t_2)}{\sigma_0^2} = -\frac{f(-t_1) f(-t_2)}{w(0)} + \frac{f'(-t_1) f'(-t_2)}{-w''(0)}\left(1-\frac{\pi}{2}\right) + 
\delta(t_2 - t_1)
\label{C_t1_t2}
\end{equation}
The eigenfunctions $\psi_i(t)$ of the covariance operator
and their corresponding eigenvalues $\lambda_i$ are determined by the equation
\begin{equation}
\int\,C(t_1,t_2) \psi_i(t_2) {\rm d}t_2 = \lambda_i \psi_i(t_1)
\end{equation}
From Eq.~(\ref{C_t1_t2}) we see that there are only two eigenfunctions $\psi_{1,2}(\Delta t)$ with eigenvalues
that  differ from $\sigma_0^2$. (The significance of these eigenfunctions is that the variance of the stimulus's
projection on $\psi_{1,2}$, when
conditioned on the occurrence of a spike at $t = 0$, is different from its nominal value
of $\sigma_0^2$.) 
The first eigenfunction $\psi_1(\Delta t)$ is proportional to $f(-\Delta t)$.
Because the generating potential must be equal to the threshold at
$t = 0$, when a spike is produced,  the variance of the stimulus's
projection on $f$ must be zero: Indeed, $\lambda_1 = 0$. The second eigenfunction
 $\psi_2(\Delta t) \propto f'(-\Delta t)$, and $\lambda_2 = (2-\pi/2)\sigma_0^2$.

\section{Spike correlations}

We consider two model neurons that may differ in their generating potentials $g_{1,2}$
and in their thresholds $\theta_{1,2}$. The generating potentials
 $g_{1,2}$ are assumed to be jointly Gaussian and stationary, characterized by
correlation functions:
\begin{eqnarray}
w_{ij}(\Delta t) = \left<g_{i}(0) g_{j}(\Delta t) \right>
\end{eqnarray}

The stationary nature of $g_{1,2}$ implies that $w_{11}(-\Delta t) = w_{11}(\Delta t)$ and that,
similarly, $w_{22}(-\Delta t) = w_{22}(\Delta t)$.
For the cross-correlation functions it only implies that $w_{12}(-\Delta t) = w_{21}(\Delta t)$. 
In the rest of this appendix we assume that, in addition,
\begin{equation}
w_{12}(\Delta t) = w_{21}(\Delta t)
\label{eq:assumption}
\end{equation}
which is correct throughout Sec.~V.
In this case the expressions for the joint probability distribution and for the spike correlation function
simplify considerably. 

We first need to evaluate the joint probability distribution for $g_1(0) = \theta_1$, $\dot{g_1}(0) = q_1$, $g_2(\Delta t) = \theta_2$, and $\dot{g_2}(\Delta t) = q_2$. 
This is given by
\begin{equation}
p(\theta_1, q_1; \theta_2; q_2) = \frac{1}{Z}\,
{\rm exp}\left[-\frac{1}{2}{\bf \zeta}^T A^{-1} {\bf \zeta}\right]
\label{eq:p_tqtq}
\end{equation}
where $\zeta^T = (\theta_1, q_1, \theta_2, q_2)$, $Z = \left[(2\pi)^n {\rm det}A\right]^{1/2}$, and
\begin{equation}
A = \left[
\begin{array}{cccc}
w_{11}(0) & 0 & w_{12}(\Delta t) & w_{12}'(\Delta t) \\
0 & -w_{11}''(0) & -w_{12}'(\Delta t) & -w_{12}''(\Delta t) \\
w_{12}(\Delta t) & -w_{12}'(\Delta t) & w_{22}(0) & 0 \\
w_{12}'(\Delta t) & -w_{12}''(\Delta t) & 0 & -w_{22}''(0)
\end{array}
\right]
\label{eq:A4_gen}
\end{equation}

Since, in calculating the spike correlation function, $\theta_1$ and $\theta_2$ are kept fixed, it is useful to re-express this quantity as a quadratic function of $q_1$ and $q_2$ alone. We know that
\begin{equation}
\int {\rm d} q_1 \int {\rm d}q_2\,p(\theta_1, q_1; \theta_2, q_2) = p(\theta_1; \theta_2)
\end{equation}
where $p(\theta_1; \theta_2)$ is the joint probability distribution of $g_1(0) = \theta_1$, $g_2(\Delta t) = \theta_2$,
\begin{equation}
p(\theta_1; \theta_2)  =  \frac{1}{2\pi D} 
 {\rm exp}\left(
-\frac{w_{11}(0)\theta_1^2 + w_{22}(0)\theta_2^2 - 2w_{12}(\Delta t) \theta_1 \theta_2}
{2D^2}
\right),
\label{eq:p_tt_gen}
\end{equation}
and
\begin{equation}
D = \left[ w_{11}(0)w_{22}(0) - w_{12}^2(\Delta t)\right]^{1/2}.
\end{equation}
Hence we can rewrite $p(\theta_1, q_1; \theta_2, q_2)$ as
\begin{equation}
p  =  p(\theta_1; \theta_2) 
 \times  \frac{\left({\rm det}\,m^{-1}\right)^{1/2}}{2\pi} 
{\rm exp}\left[-\frac{1}{2}\left({\bf q} - {\bf q}_0\right)^Tm^{-1}\left({\bf q}-{\bf q}_0\right)\right]
\nonumber
\end{equation}

The matrix $m$ and ${\bf q}_0$ are found by collecting the quadratic and linear terms in 
$q_1$, $q_2$ in Eq.~(\ref{eq:p_tqtq}), 
\begin{equation}
m  =  \left[\begin{array}{cc}
-w_{11}''(0) & -w_{12}''(\Delta t) \\
-w_{12}''(\Delta t) & -w_{22}''(0)
\end{array}\right]
 - 
\frac{\left[w_{12}'(\Delta t)\right]^2}{w_{11}(0)w_{22}(0)-w_{12}^2(\Delta t)}
\left[\begin{array}{cc}
w_{11}(0) & w_{12}(\Delta t) \\
w_{12}(\Delta t) & w_{22}(0)
\end{array}\right],
\label{eq:m_gen} 
\end{equation}
and
\begin{equation}
{\bf q}_0 = \frac{w_{12}'(\Delta t)}{w_{11}(0)w_22(0)-w_{12}^2(\Delta t)}\left[\begin{array}{c}
w_{12}(\Delta t) \theta_1 - w_{11}(0) \theta_2 \\
w_{22}(0) \theta_1 - w_{12}(\Delta t) \theta_2
\end{array}\right]
\label{eq:q0_gen}
\end{equation}
Finally, we can write the spike correlation function as
\begin{equation}
c(\Delta t)  = p(\theta_1;\theta_2) \cdot I
\label{eq:c_p_I}
\end{equation}
where
\begin{equation}
I = \frac{\left({\rm det}\,m^{-1}\right)^{1/2}}{2\pi} 
\int_0^\infty {\rm d}q_1 \int_0^\infty {\rm d}q_2\,q_1 \cdot q_2 
\cdot {\rm exp}\left[-\frac{1}{2}\left({\bf q} - {\bf q}_0\right)^Tm^{-1}\left({\bf q}-{\bf q}_0\right)\right]
\label{eq:Appendix_I}
\end{equation}

\subsubsection*{Nearly independent inputs}

When the cross-correlation $w_{12}(\Delta t)$ is small compared to the auto-correlations $w_{11}(0)$, $w_{22}(0)$, 
$I$ and $p$ can be expanded in powers of $w_{12}(\Delta t)$ and its derivatives. If
$w_{12}(\Delta t) = 0$ the matrix $m$ is diagonal,  $p(\theta_1, q_1; \theta_2, q_2) = p(\theta_1,q_1)\cdot p(\theta_2,q_2)$,
and $c (\Delta t) = r_1 r_2$. To first order in $w_{12}(\Delta t)$,
\begin{equation}
I \simeq I_0 -\frac{1}{4} w_{12}''(\Delta t) + 
\frac{1}{2 \sqrt{2\pi}}
\left\{
\frac{\left[-w_{11}''(0)\right]^{1/2}}{w_{22}(0)} \theta_1
- \frac{\left[-w_{22}''(0)\right]^{1/2}}{w_{11}(0)} \theta_2
\right\}
w_{12}'(\Delta t)
+ \cdots
\label{eq:I_large_t}
\end{equation}
where $I_0 = \left[w_{11}''(0) w_{22}''(0)\right]^{1/2}/(2\pi)$. By similarly expanding
$p(\theta_1;\theta_2)$ we find that
\begin{eqnarray}
 \frac{c(\Delta t)-r_1 r_2}{r_1 r_2} & \simeq &
 \frac{\theta_1 \theta_2}{w_{11}(0) w_{22}(0)}  w_{12}(\Delta t)
 -  \frac{\pi}{2\left[w''_{11}(0) w''_{22}(0)\right]^{1/2}}w''_{12}(\Delta t)
\nonumber \\
& + &  \sqrt{\frac{\pi}{2}} \left\{
\frac{\theta_1}{\left[-w''_{22}(0)\right]^{1/2} w_{22}(0)} - \frac{\theta_2}{\left[-w''_{11}(0)\right]^{1/2} w_{11}(0)}
\right\}
w'_{12}(\Delta t) 
\label{eq:weak_correlation} 
\end{eqnarray}

\subsubsection*{Identical inputs}

Equation (\ref{eq:Appendix_I}) can be further simplified in the case where $g_1$ and $g_2$
are identical. In this case $w_{11} = w_{12} = w_{21} = w_{22} \equiv w$.
The spike correlation function involves five parameters:
$w$ and its second derivative at zero, $w$, its derivative, and its second derivative at $\Delta t$. 
In Eq.~(\ref{eq:Appendix_I}), $m$ and ${\bf q}_0$, Eqs.~(\ref{eq:m_gen})--(\ref{eq:q0_gen}), 
are then 
\begin{equation}
m  =  \left[\begin{array}{cc}
-w''(0) & -w''(\Delta t) \\
-w''(\Delta t) & -w''(0)
\end{array}\right] 
 -  \frac{\left[w'(\Delta t)\right]^2}{w^2(0)-w^2(\Delta t)}
\left[\begin{array}{cc}
w(0) & w(\Delta t) \\
w(\Delta t) & w(0)
\end{array}\right],
\label{eq:m}
\end{equation}
\begin{equation}
{\bf q}_0 = \frac{w'(\Delta t)}{w^2(0)-w^2(\Delta t)}\left[\begin{array}{c}
w(\Delta t) \theta_1 - w(0) \theta_2 \\
w(0) \theta_1 - w(\Delta t) \theta_2
\end{array}\right],
\label{eq:q0}
\end{equation}
and 
\begin{equation}
p(\theta_1; \theta_2)  =  \frac{1}{2\pi \left[ w(0)^2 - w(\Delta t)^2\right]^{1/2}} 
 \cdot {\rm exp}\left\{
-\frac{w(0)(\theta_1^2 + \theta_2^2) - 2w(\Delta t) \theta_1 \theta_2}
{2\left[w(0)^2 - w(\Delta t)^2\right]}
\right\}
\label{eq:p_tt} 
\end{equation}
Note that from symmetry under time reversal we must have
\begin{equation}
p(\theta_1, q_1; \theta_2, q_2) = p(\theta_2,-q_2; \theta_1,-q_1)
\label{eq:p_symmetry}
\end{equation}
This symmetry is reflected in the fact that $({\bf q}_0)_{1,2}(\theta_1,\theta_2) = -({\bf q}_0)_{2,1}(\theta_2,\theta_1)$.

Because $m_{11} = m_{22}$, the eigenvectors of $m$ are
\begin{equation}
{\bf v}_\pm = \frac{1}{\sqrt{2}} \left(\begin{array}{c}1 \\ \pm 1\end{array}\right)
\end{equation}
and the corresponding eigenvalues are
\begin{equation}
m_\pm  =  -w''(0) \mp w''(\Delta t)
 +  \frac{\left[w'(\Delta t)\right]^2}{w^2(0)-w^2(\Delta t)}
\left[-w(0) \mp w(\Delta t)\right]
\end{equation}
It is useful to rewrite $I$ in the coordinates in which $m$ is diagonal
\begin{equation}
u_\pm = \frac{1}{\sqrt{2}}(q_1 \pm q_2)
\label{eq:u_pm_def}
\end{equation}
in terms of which
\begin{eqnarray}
I & = & \frac{1}{2}\int_0^\infty {\rm d}u_+ \int_{-u_+}^{u_+} {\rm d}u_-\,(u_+^2 -u_-^2) 
\label{eq:I_u_integral}
\\
& \times &
\frac{1}{2\pi m_+^{1/2} m_-^{1/2}} 
{\rm exp}\left[-\frac{(u_+-u_{0,+})^2}{2m_+} - \frac{(u_--u_{0,-})^2}{2m_-}\right]
\nonumber
\end{eqnarray}
where
\begin{equation}
u_{0,\pm} = -\frac{1}{\sqrt{2}} \frac{w'(\Delta t)}{w(0) \mp w(\Delta t)} (\theta_2\mp\theta_1)
\label{eq:u_0}
\end{equation}
The inner integral can be expressed using the Gauss error function, yielding an expression for $I$ that involves a single integral,
\begin{eqnarray}
I & = & \int_0^{\infty} {\rm d}u_+ \, \frac{1}{4\pi m_+^{1/2}} {\rm exp}\left[-\frac
{(u_+-u_{0,+})^2}{2m_+} \right]
\\
& \times & \left\{ 
m_-^{1/2} (u_+ -u_{0,-}) {\rm exp} \left[ -\frac{(u_++u_{0,-})^2}{2m_-}\right] \right.
\nonumber \\
& & + 
m_-^{1/2} (u_+ +u_{0,-}) {\rm exp} \left[ -\frac{(u_+-u_{0,-})^2}{2m_-}\right] 
\nonumber \\
& &
\left.
+ \sqrt{\frac{\pi}{2}} \left(m_-+u_{0,-}^2-u_+^2\right)
\left[
{\rm erf}\left(\frac{u_{0,-}-u_+}{\sqrt{2m_-}}\right)
-{\rm erf}\left(\frac{u_{0,-}+u_+}{\sqrt{2m_-}}\right)
\right] 
\right\} \nonumber
\label{eq:I_long}
\end{eqnarray}

\subsubsection*{The limit $\Delta t \rightarrow \infty$}

Assuming that $w$ and all its derivatives at $t = \Delta t$ tend to zero when $\Delta t \rightarrow \infty$,
the probability distributions at $t = 0$ and at $t = \Delta t$ decouple, and to leading order
$c(\Delta t) \rightarrow r_1 r_2$.

To evaluate the deviation from independent spiking, we can use equation (\ref{eq:weak_correlation}) in the case
where $w_{11} = w_{22} = w_{12} = w_{21} \equiv w$:
\begin{eqnarray}
 \frac{c(\Delta t)- r_1 r_2}{r_1 r_2} & \simeq &
 \frac{\theta_1\theta_2}{w(0)^2} w(\Delta t)
+ \frac{\pi}{2w''(0)}w''(\Delta t)  
\label{eq:spike_corr_large_t} \\
 & + & \frac{1}{w(0)}\left[\frac{\pi}{-2w''(0)}\right]^{1/2}
 (\theta_1-\theta_2) w'(\Delta t) 
 \nonumber
\end{eqnarray}
The third term in this equation is antisymmetric in $\Delta t$ and in $\theta_1-\theta_2$. Because $w'(\Delta t)$ is typically negative for positive $\Delta t$, this term represents a small preference for the neuron with the higher threshold to spike after the neuron with the lower threshold.

\section{The limit $\Delta t \rightarrow 0$}

We consider here the limit $\Delta t \rightarrow 0$ in the case of identical inputs. In this limit the matrix $m$ becomes singular,
requiring particular analysis in order to evaluate the correlation matrix.

The argument of the exponential in Eq.~(\ref{eq:I_def}) 
is maximal when ${\bf q} = {\bf q}_0$ [Eq.~(\ref{eq:q0})] 
which approaches, when $\Delta t \rightarrow 0$, 
\begin{equation}
(q_0)_{1,2} \rightarrow \frac{\theta_2-\theta_1}{\Delta t}
\end{equation}
This is the derivative of $g(t)$ if it follows a linear trajectory from $g(0) = \theta_1$ to $g(\Delta t) = \theta_2$. However, this maximum is within the integration range in Eq.~(\ref{eq:I_def}) 
only if $(\theta_2-\theta_1) / \Delta t$ is positive, \textit{i.e.}, only if the neuron with the higher threshold spikes after the 
neuron with the lower threshold. Accordingly, the behavior of $c(\Delta t)$ is 
substantially different for positive and negative $\Delta t$. 

For simplicity of the notation in this appendix, we assume that $\Delta t > 0$ and that $\theta_2 - \theta_1$ may be either positive,
negative, or zero. To relate this to the presentation in Sec.~V, where we assumed that $\theta_2 - \theta_1$ is positive, recall that $c(-\Delta t)$ is the same as $c(\Delta t)$ with the
two thresholds $\theta_1,\theta_2$ exchanged.

To evaluate the behavior of the integral $I$ at small $\Delta t$ we need to expand the expressions for
$m_+$, $m_-$, $u_+$, and $u_-$ in this limit, where we use the representation of $I$ in 
Eqs.~(\ref{eq:I_u_integral})--(\ref{eq:u_0}).
The leading terms in these expansions are shown in Table~1 in two cases: (i) $W_3 \neq 0$. (ii) All $W_i$ with odd index vanish.

\begin{table}[h]
\begin{center}
\renewcommand{\arraystretch}{2.5}
\begin{tabular}{|c||c|c|}
\hline
& $W_3 \neq 0$ & $W_3 = W_5 = 0$ \\
\hline
\hline
$m_+ $ &  $\displaystyle \frac{W_3}{3} \Delta t$ & $\displaystyle \frac{1}{72}\left(\frac{W_4^2}{W_2}-W_6\right)(\Delta t)^4$ \\
\hline
$m_-$ & $W_3\Delta t$ & $\displaystyle \frac{1}{2}\left(-\frac{W_2^2}{W_0}+W_4\right) (\Delta t)^2$ \\
\hline
$\sqrt{2}u_+ $ & \multicolumn{2}{|c|}{$\displaystyle (\theta_2 - \theta_1) \frac{2}{\Delta t}$}  \\
\hline
$\sqrt{2}u_-$ & \multicolumn{2}{|c|}{$\displaystyle -(\theta_1 + \theta_2) \frac{W_2}{2 W_0} \Delta t$} \\
\hline
\end{tabular}
\end{center}
\caption{Leading order terms in the expansion of $m_\pm$ and $u_\pm$ for small $\Delta t$. 
If $\theta_1 = \theta_2$, $u_+ = 0$ to all orders.}
\end{table}

\subsubsection*{The case $\theta_2 - \theta_1 > 0$}

This case is most simply treated in the original ({\bf q}) coordinates. As $\Delta t \rightarrow 0$
$(q_{0})_{1,2} \sim (\Delta t)^{-1}$ whereas the standard deviation of the Gaussian 
in Eq.~(\ref{eq:I_def}) tends to zero as $\sqrt{\Delta t}$ in both of the principal directions
(or faster if $W_3 = 0$). We can therefore
replace the lower integration limits by $-\infty$ with an exponentially small error. Furthermore,
the integral in Eq.~(\ref{eq:I_def}) becomes dominated by the 
peak of the Gaussian, as it becomes sharper with the decrease of $\Delta t$.

To leading order in $\Delta t$  the integral is thus simply
\begin{equation}
I \simeq (q_0)_1 \cdot (q_0)_2  = \frac{(\theta_2-\theta_1)^2}{\Delta t^2} 
\label{eq:I_small_t_pos_ap}
\end{equation}

To evaluate $p(\theta_1;\theta_2)$ we note that, in Eq.~(\ref{eq:p_tt}),
\begin{equation}
w^2(0)-w^2(\Delta t) \simeq -W_0 W_2 (\Delta t)^2 
\end{equation}
and by expanding the argument of the exponential we obtain
\begin{equation}
p(\theta_1;\theta_2) \simeq  \frac{1}{2\pi (-W_0 W_2)^{1/2}|\Delta t|} 
 \times 
{\rm exp}\left[\frac{1}{2W_2}\left(\frac{\theta_2-\theta_1}{\Delta t}\right)^2+ \frac{\gamma_1}{\Delta t} + \gamma_0\right],
\label{eq:p_tt_small_t} 
\end{equation}
where
\begin{equation}
\gamma_0 =  \frac{1}{72 W_2^3}(4 W_3^2-3 W_2 W_4)(\theta_2-\theta_1)^2 -
\frac{1}{8 W_0} (\theta_1+\theta_2)^2.
\label{eq:gamma_0}
\end{equation}
and
\begin{equation}
\gamma_1  =  -\frac{W_3}{6 W_2^2}(\theta_2-\theta_1)^2.
\label{eq:gamma1}
\end{equation}
The leading contribution in Eq.~(\ref{eq:p_tt_small_t}) is proportional to the probability density to have a derivative equal to $(\theta_2-\theta_1)/\Delta t$.
Combining Eqs.~(\ref{eq:I_small_t_pos_ap}) and (\ref{eq:p_tt_small_t}) 
we obtain the small $\Delta t$ behavior of $c(\Delta t)$, Eq.~(\ref{eq:p_tt_small_t_approx}).

When $\theta_2 - \theta_1 \rightarrow 0$ the peak in the spike correlation function gradually becomes the delta-function contribution to the spike auto-correlation function, Eq.~(\ref{eq:delta_contribution}):
For sufficiently small $\theta_2 - \theta_1$, and for $\Delta t \lesssim -W_2/W_3$, the spike correlation function is approximately
\begin{equation}
c(\Delta t)  \simeq 
\frac{(\theta_2-\theta_1)^2}{2\pi (-W_0 W_2)^{1/2}\Delta t^3
}
 \times  {\rm exp}\left[-\frac{\theta^2}{2W_0}
+ \frac{1}{2W_2}\left(\frac{\theta_2-\theta_1}{\Delta t}\right)^2
\right].
\end{equation}
The integral of this function from $0$ to infinity is equal to $r$, whereas the width of the peak and its position both tend to zero when $\theta_2 - \theta_1\rightarrow 0$. Hence the small $\Delta t$ behavior of the spike correlation function approaches, when $\theta_2 - \theta_1 \rightarrow 0$, 
\begin{equation}
 r \delta (\Delta t).
 \end{equation}

\subsubsection*{The case $\theta_2 = \theta_1$}

The maximum of the Gaussian in Eq.~(\ref{eq:I_u_integral}) is at $u_+ = 0$ and $u_- \sim \theta \Delta t$ (where $\theta \equiv \theta_1 = \theta_2$,) which lies outside the range of integration
[The (+,+) quadrant in the {\bf q} coordinates.] However, if $W_3 \neq 0$ the distance from the (+,+) quadrant is small compared to the standard deviation of the Gaussian ($\sim \sqrt{\Delta t}$). As a result, the integral in (\ref{eq:I_u_integral}) can be treated as if $u_{0,\pm} = 0$. The situation is different if $W_3 = 0$, in which case the distance of ${\bf q}$ from the origin and the standard deviation in the $u_-$ direction are of the same order of magnitude. 

\noindent (\textit{i}) $W_3 \neq 0$. In this case we have, to leading order in $\Delta t$,
\begin{eqnarray}
I & \simeq & \frac{1}{4\pi m_+^{1/2} m_-^{1/2}} \int_0^\infty {\rm d}u_+ \int_{-u_+}^{u_+} {\rm d}u_+\,(u_+^2-u_-^2)
 {\rm exp} \left(-\frac{u_+^2}{2m_+} - \frac{u_-^2}{2m_-}
\right)
\nonumber \\
& = & \frac{1}{2\pi} \left[(m_+ m_-)^{1/2}  + (m_+-m_-) {\rm arctan} \sqrt{\frac{m_+}{m_-}} \right]
\nonumber \\
& \simeq &
\frac{W_3 \Delta t}{2\pi \sqrt{3}}\left[1-\frac{\pi}{3\sqrt{3}}\right].
\label{eq:approx_I_1}
\end{eqnarray}
For small $\Delta t$
\begin{equation}
p(\theta;\theta) \simeq \frac{1}{2\pi\sqrt{-W_0 W_2} \Delta t} {\rm exp}\left(-\frac{\theta^2}{2W_0}
\right)
\label{eq:p_tt_same}
\end{equation}
Hence the spike correlation function tends to a finite value when $\Delta t \rightarrow 0$,
Eq.~(\ref{eq:c_deltat_auto_w3}). 

\noindent {({\textit ii}) $W_3 = W_5 = 0$. To treat this case we rescale $u_\pm$ by the leading dependence on $\Delta t$ in the expansion of  $(m_\pm)^{1/2}$  (this procedure can also be applied in the case $W_3 \neq 0$, in order to derive Eq.~(\ref{eq:approx_I_1}) in a more formal manner). We use the following notation,
\begin{eqnarray}
m_+ & \simeq & a_+ (\Delta t)^4 \ \ \ \ \ ; \ \ \ \ \ m_-  \simeq  a_- (\Delta t)^2 \nonumber \\
u_+  & = & \tilde{u}_+ \cdot (\Delta t)^2 \ \ \ \ \ ; \ \ \ \ \ 
u_-  =  \tilde{u}_- \cdot \Delta t \ \ \ \ \ ; \ \ \ \ \  
u_{-,0}  \simeq \alpha \Delta t 
\end{eqnarray}
where 
\begin{eqnarray}
a_+ & = & \frac{1}{72}\left(\frac{W_4^2}{W_2}-W_6\right)  \ \ ;\ \ \ 
a_- = \frac{1}{2}\left(-\frac{W_2^2}{W_0}+W_4\right) 
\nonumber \\
\alpha & = & -\frac{W_2}{\sqrt{2}W_0} \theta
\end{eqnarray}
and obtain
\begin{eqnarray}
I  & \simeq & \frac{1}{4\pi (a_+ a_-)^{1/2}} \int_0^\infty {\rm d}\tilde{u}_+ \int_{-\tilde{u}_+ \Delta t}^{\tilde{u}_+ \Delta t}
{\rm d}\tilde{u}_- \,  \left[ (\Delta t)^4 \tilde{u}_+^2 - (\Delta t)^2 \tilde{u}_-^2\right] 
\nonumber \\
& & \times {\rm exp}
\left[ -\frac{\tilde{u}_+^2}{2a_+} - \frac{(\tilde{u}_--\alpha)^2}{2a_-}\right]
\nonumber \\
& \simeq 
& \frac{2}{3\pi} (\Delta t)^5 \sqrt{\frac{a_+^3}{a_-}}{\rm exp}\left(-\frac{\alpha^2}{2a_-}\right)
\label{eq:I_tt_same_approx}
\end{eqnarray}
Using Eq.~(\ref{eq:p_tt_same}) 
we see that, in contrast to the case $W_3\neq 0$, the spike correlation function tends to zero  when $\Delta t \rightarrow 0$,
\begin{equation}
c(\Delta t) \sim (\Delta t)^4
\end{equation}
\subsubsection*{The case $\theta_2 - \theta_1 < 0$}
In this case, as in the case $\theta_2 > \theta_1$, $(q_0)_{1,2} \sim \Delta t^{-1}$, but now ${\bf q}_0$ is outside the integration region, far from it on the scale of the Gaussian's standard deviation.  
We denote
\begin{equation}
u_{+,0} \simeq -\frac{\alpha}{\Delta t}
\end{equation}
where $\alpha = \sqrt{2}(\theta_1-\theta_2) > 0$. It is then seen from Eq.~(\ref{eq:I_u_integral}) that $I$ decays
to zero exponentially, with a leading contribution of the form
\begin{equation}
{\rm log}(I) \simeq -\frac{\alpha^2}{2m_+ (\Delta t)^2} = -\frac{1}{m_+}\left(\frac{\theta_1-\theta_2}{\Delta t}\right)^2
\end{equation}
which corresponds, if $W_3 \neq 0$ to
\begin{equation}
{\rm log}(I) \simeq -\frac{3(\theta_1-\theta_2)^2}{W_3(\Delta t)^3} + \cdots
\end{equation}
and, if $W_3 = W_5 = 0$, to
\begin{equation}
{\rm log}(I) \simeq -\frac{(\theta_1-\theta_2)^2}{a_+(\Delta t)^6} + \cdots
\end{equation}
In contrast to $I$ in the case $\theta_2 - \theta_1 > 0$, which diverges when $\Delta t \rightarrow 0$,
in the case $\theta_2 - \theta_1 < 0$ $I$ strongly decays to zero. 
A more precise treatment of this integral requires evaluation of additional terms in the expansion shown in Table~1. 

The probability $p(\theta_1; \theta_2)$ is the same as in the case $\theta_2 - \theta_1 > 0$. Like $I$, $p$ decays exponentially when ${\Delta t} \rightarrow 0$, but the leading power of $(\Delta t)^{-1}$ inside the exponential is smaller, so it becomes relevant only if
${\rm log}I$ is expanded beyond the leading order.

\section{Parameterization of the LIF model}

The notation in  equations (\ref{eq:IAF1})--(\ref{eq:IAF2}) was chosen for mathematical simplicity. 
To see how the parameters in these equations are related to biophysical properties of the neuron,
we start with a standard equation for the membrane potential dynamics of a LIF neuron
\cite{DayanAbbott},
\begin{equation}
\tau_1 \frac{{\rm d}v}{{\rm d t}} = E_0 -v + R_m I_c
\end{equation}
where $E_0$ is the resting potential of the cell, $R_m$ is the membrane
resistance, and $I_c$ is the input current. Further, the potential $v$ resets to
$v_r$ whenever it reaches a threshold $v_\theta$. 

To reparametrize this equation
as in Eq.~(\ref{eq:IAF1}), we first define
\begin{equation}
I = R_m\left(I_c - \left<I_c\right>\right)
\label{I_Ic}
\end{equation}
where $\left< I_c \right>$ is the mean (baseline) value of the fluctuating current $I_c$.
Due to the multiplication by $R_m$, $I$ has the same
units as the potential $u$. Second, we shift the potential $v$ by defining:
\begin{equation}
u = v-E_0 - R_m \left<I_c\right>
\end{equation}
With these definitions we obtain Eq.~(\ref{eq:IAF1}), repeated here,
\begin{equation}
\tau_1 \frac{{\rm d}u}{{\rm d}t} = -u + I
\end{equation}
where $I$ has zero mean, and $\theta - u_r = v_\theta - v_r$.

The properties of the spike train depend
on the ratio between $\theta - u_r$ and the standard deviation $\sigma$.
The following example demonstrates what is a reasonable range of values for this ratio.
The standard deviation of $I$ is equal to $\sqrt{2}\sigma$ (assuming that $\tau_1 = \tau_2$).
Suppose that the standard deviation of the current $I_c$ is $0.1$\,nA.
Assuming that $R_m = 100$\,M$\Omega$ (which corresponds to a neuron with surface area $10^{-4}$\,cm$^2$ and
membrane time constant $\tau_1 = 10$\,ms), and that the difference between
the threshold membrane potential and the reset potential is 10\,mV we get,
using Eq.~(\ref{I_Ic}), $\theta-u_r \simeq 0.7\sigma$. 

The
ratio $(\theta-u_r)/\sigma$ can vary considerably depending on the biophysical parameters
of the cell and, more importantly, depending on the standard deviation of the fluctuating
current. A smaller standard deviation of the fluctuating current corresponds to a larger value of
$(\theta-u_r)/\sigma$.

\section{Correlations in the LN model}

In the LN model the spike times are an inhomogeneous Poisson process, with a rate given by
\begin{equation}
r(t) = \phi \left[(f \circ s) (t) \right]
\label{eq:ln_model}
\end{equation}
where $\phi$ may be regarded as the transfer function of the neuron, if we think of $s$ as
a direct input to the neuron (more generally, we may think of
Eq.~(\ref{eq:ln_model}) as a phenomenological relation between stimulus and firing rate.)

If there was no nonlinearity [$\phi(x) = x$] we would have $\left<r(t)r(t')\right> \sim  w(t-t')$,
where $w$ is the auto-correlation function of $f\circ s$. More generally,
we need to take into account the particular form of $\phi$. We focus here on the case where
two neurons, labeled $1$ and $2$, receive the same input and have the same filter, but possibly
differ in the nonlinear function $\phi$. As in previous sections we assume that the
stimulus $s$ is Gaussian (and so is $g = f\circ s$).  We then have,
\begin{equation}
c(\Delta t) = \int_{-\infty}^{\infty} {\rm d}g_1 
\int_{-\infty}^\infty {\rm d}g_2 \, \phi_1(g_1) \phi_2(g_2) p(g_1;g_2)
\end{equation}
where $p(g_1;g_2)$ is given by Eq.~(\ref{eq:p_tt}). 
The spike correlation function thus depends on $w(0)$
and on $w(\Delta t)$: in comparison, in the threshold crossing model it depends also 
on the first and second derivatives of $w$ at these two points.

For example, if
\begin{equation}
\phi_{1,2}(g) = \phi_0 \left\{
\begin{array}{lll}
0 & , & g < \theta_{1,2} \\
g-\theta_{1,2} & , & g > \theta_{1,2}
\end{array}
\right.
\end{equation}
the spike correlation function is
\begin{equation}
c(\Delta t) = \phi_0^2 \int_{\theta_1}^\infty {\rm d}g_1 
\int_{\theta_2}^\infty {\rm d}g_2 (g_1-\theta_1)(g_2-\theta_2)p(g_1,g_2)
\label{eq:c_LN_general}
\end{equation}
Note that the spike correlation function in the LN model
is symmetric in $\Delta t \rightarrow -\Delta t$ despite the different
offsets for the two neurons. 

In the special case where $\theta_1 = \theta_2 = 0$, the spike (auto-)correlation function can be written in a relatively 
simple form,
\begin{equation}
c(\Delta t)  =  \frac{1}{2\pi} \sqrt{w^2(0)-w^2(\Delta t)} +  \frac{w(\Delta t)}{4}
 +  \frac{|w(\Delta t)|}{2\pi} {\rm arctan}\frac{|w(\Delta t)|}{\sqrt{w^2(0)-w^2(\Delta t)}}
\end{equation}

For any two thresholds, the behavior for large $|\Delta t|$ can be found from the expansion of $p(g_1; g_2)$, from which we obtain
\begin{equation}
c(\Delta t)   \simeq  r_{0,1} r_{0,2} 
 +  \frac{\phi_0^2}{4} w(\Delta t) 
{\rm erfc}\left(\frac{\theta_1}{\sqrt{2w(0)}}\right)
{\rm erfc}\left(\frac{\theta_2}{\sqrt{2w(0)}}\right)
\end{equation}
where 
\begin{equation}
r_{0,i} = \sqrt{\frac{w(0)}{2\pi}}{\rm exp}\left(-\frac{\theta_i^2}{2w(0)}\right)
-\frac{\theta_i}{2} {\rm erfc}\left(\frac{\theta_i}{\sqrt{2w(0)}}\right)
\end{equation}
are the firing rates of the two neurons and erfc is the complementary error function.

For a more direct comparison with the threshold crossing model it is instructive to consider a transfer function that is concentrated at a particular value $\theta$: $\phi(x) = \phi_0 \delta(x-\theta)$. We then find that for large $|\Delta t|$,
\begin{equation}
\frac{c(\Delta t)- r_{1,0}r_{2,0}}{r_{1,0} r_{2,0}} \simeq 
\frac{w(\Delta t)}{w^2(0)} \theta_1 \theta_2
\label{eq:c_LN_delta}
\end{equation}
which is the same as the first term in Eq.~(\ref{eq:spike_corr_large_t}). The firing rate in the LN model, appearing in the
left hand side of Eq.~(\ref{eq:c_LN_delta}), is given by
\begin{equation}
r_{0,i} = \frac{\phi_0}{\sqrt{2\pi w(0)}} {\rm exp}\left[-\frac{\theta_i^2}{2w(0)}\right]
\end{equation}
%

\begin{thebibliography}{}

\bibitem[\protect\citeauthoryear{%
Baccus%
\ \BBA{} Meister%
}{%
Baccus%
\ \BBA{} Meister%
}{%
{\protect\APACyear{2002}}%
}]{%
Baccus2002}%
\APACinsertmetastar{%
Baccus2002}%
Baccus, S\BPBI A.%
\BCBT{}\ \BBA{} Meister, M.%
%
\unskip\
\newblock
\APACrefYearMonthDay{2002}{}{}.
\newblock
\BBOQ{}\APACrefatitle{Fast and slow contrast adaptation in retinal
  circuitry}{Fast and slow contrast adaptation in retinal circuitry}.\BBCQ{}
\newblock
\APACjournalVolNumPages{Neuron}{36}{}{909--919}.
\PrintBackRefs{\CurrentBib}

\bibitem[\protect\citeauthoryear{%
Berry%
\ \BBA{} Meister%
}{%
Berry%
\ \BBA{} Meister%
}{%
{\protect\APACyear{1998}}%
{\protect\APACexlab{{\protect\BCnt{1}}}}}]{%
Berry1999}%
\APACinsertmetastar{%
Berry1999}%
Berry, M\BPBI J.%
\BCBT{}\ \BBA{} Meister, M.%
%
\unskip\
\newblock
\APACrefYearMonthDay{1998{\protect\BCnt{1}}}{}{}.
\newblock
\BBOQ{}\APACrefatitle{Refractoriness and neural precision}{Refractoriness and
  neural precision}.\BBCQ{}
\newblock
\APACjournalVolNumPages{Journal of Neuroscience}{18}{}{2200--2211}.
\PrintBackRefs{\CurrentBib}

\bibitem[\protect\citeauthoryear{%
Berry%
\ \BBA{} Meister%
}{%
Berry%
\ \BBA{} Meister%
}{%
{\protect\APACyear{1998}}%
{\protect\APACexlab{{\protect\BCnt{2}}}}}]{%
MeisterBerry1998}%
\APACinsertmetastar{%
MeisterBerry1998}%
Berry, M\BPBI J.%
\BCBT{}\ \BBA{} Meister, M.%
%
\unskip\
\newblock
\APACrefYearMonthDay{1998{\protect\BCnt{2}}}{}{}.
\newblock
\BBOQ{}\APACrefatitle{Refractoriness and neural prediction}{Refractoriness and
  neural prediction}.\BBCQ{}
\newblock
\APACjournalVolNumPages{Journal of Neuroscience}{18}{}{2200--2211}.
\PrintBackRefs{\CurrentBib}

\bibitem[\protect\citeauthoryear{%
Berry%
, Warland%
\BCBL{}\ \BBA{} M.%
}{%
Berry%
\ \protect\BOthers{.}}{%
{\protect\APACyear{1997}}%
}]{%
Berry1997}%
\APACinsertmetastar{%
Berry1997}%
Berry, M\BPBI J.%
, Warland, D\BPBI K.%
\BCBL{}\ \BBA{} M., M.%
%
\unskip\
\newblock
\APACrefYearMonthDay{1997}{}{}.
\newblock
\BBOQ{}\APACrefatitle{The structure and precision of retinal spike trains}{The
  structure and precision of retinal spike trains}.\BBCQ{}
\newblock
\APACjournalVolNumPages{Proceedings of the National Academy of Sciences of the
  USA}{94}{}{5411--5416}.
\PrintBackRefs{\CurrentBib}

\bibitem[\protect\citeauthoryear{%
Brunel%
\ \BBA{} Sergi%
}{%
Brunel%
\ \BBA{} Sergi%
}{%
{\protect\APACyear{1998}}%
}]{%
BrunelSergi1998}%
\APACinsertmetastar{%
BrunelSergi1998}%
Brunel, N.%
\BCBT{}\ \BBA{} Sergi, S.%
%
\unskip\
\newblock
\APACrefYearMonthDay{1998}{}{}.
\newblock
\BBOQ{}\APACrefatitle{Firing frequency of leaky integrate-and-firing neurons
  with synaptic current dynamics}{Firing frequency of leaky
  integrate-and-firing neurons with synaptic current dynamics}.\BBCQ{}
\newblock
\APACjournalVolNumPages{Journal of Theoretical Biology}{195}{}{87--95}.
\PrintBackRefs{\CurrentBib}

\bibitem[\protect\citeauthoryear{%
Burkitt%
}{%
Burkitt%
}{%
{\protect\APACyear{2006}}%
}]{%
Burkitt2006}%
\APACinsertmetastar{%
Burkitt2006}%
Burkitt, A\BPBI N.%
%
\unskip\
\newblock
\APACrefYearMonthDay{2006}{}{}.
\newblock
\BBOQ{}\APACrefatitle{A review of the integrate-and-fire neuron model: I.
  Homogeneous synaptic input}{A review of the integrate-and-fire neuron model:
  I. homogeneous synaptic input}.\BBCQ{}
\newblock
\APACjournalVolNumPages{Biological Cybernetics}{95}{}{1--19}.
\PrintBackRefs{\CurrentBib}

\bibitem[\protect\citeauthoryear{%
E.~Chichilnisky%
}{%
E.~Chichilnisky%
}{%
{\protect\APACyear{2001}}%
}]{%
Chichilnisky2001}%
\APACinsertmetastar{%
Chichilnisky2001}%
Chichilnisky, E.%
%
\unskip\
\newblock
\APACrefYearMonthDay{2001}{}{}.
\newblock
\BBOQ{}\APACrefatitle{A simple white noise analysis of neuronal light
  responses}{A simple white noise analysis of neuronal light responses}.\BBCQ{}
\newblock
\APACjournalVolNumPages{Network: Computation in Neural
  Systems}{12}{}{199--213}.
\PrintBackRefs{\CurrentBib}

\bibitem[\protect\citeauthoryear{%
E\BPBI J.~Chichilnisky%
\ \BBA{} Kalmar%
}{%
E\BPBI J.~Chichilnisky%
\ \BBA{} Kalmar%
}{%
{\protect\APACyear{2002}}%
}]{%
Chichilnisky2002}%
\APACinsertmetastar{%
Chichilnisky2002}%
Chichilnisky, E\BPBI J.%
\BCBT{}\ \BBA{} Kalmar, R\BPBI S.%
%
\unskip\
\newblock
\APACrefYearMonthDay{2002}{}{}.
\newblock
\BBOQ{}\APACrefatitle{Functional asymmetries in on and off ganglion cells of
  primate retina}{Functional asymmetries in on and off ganglion cells of
  primate retina}.\BBCQ{}
\newblock
\APACjournalVolNumPages{Journal of Neuroscience}{22}{}{2737--2747}.
\PrintBackRefs{\CurrentBib}

\bibitem[\protect\citeauthoryear{%
Dayan%
\ \BBA{} Abbott%
}{%
Dayan%
\ \BBA{} Abbott%
}{%
{\protect\APACyear{2001}}%
}]{%
DayanAbbott}%
\APACinsertmetastar{%
DayanAbbott}%
Dayan, P.%
\BCBT{}\ \BBA{} Abbott, L.%
%
\unskip\
\newblock
\APACrefYear{2001}.
\newblock
\APACrefbtitle{Theoretical Neuroscience}{Theoretical neuroscience}.
\newblock
\APACaddressPublisher{Cambridge, MA, USA}{The MIT Press}.
\PrintBackRefs{\CurrentBib}

\bibitem[\protect\citeauthoryear{%
Gerstner%
\ \BBA{} Kistler%
}{%
Gerstner%
\ \BBA{} Kistler%
}{%
{\protect\APACyear{2002}}%
}]{%
Gerstner}%
\APACinsertmetastar{%
Gerstner}%
Gerstner, W.%
\BCBT{}\ \BBA{} Kistler, W\BPBI M.%
%
\unskip\
\newblock
\APACrefYear{2002}.
\newblock
\APACrefbtitle{Spiking Neuron Models}{Spiking neuron models}.
\newblock
\APACaddressPublisher{Cambridge, UK}{Cambridge University Press}.
\PrintBackRefs{\CurrentBib}

\bibitem[\protect\citeauthoryear{%
Ginzburg%
\ \BBA{} Sompolinsky%
}{%
Ginzburg%
\ \BBA{} Sompolinsky%
}{%
{\protect\APACyear{1994}}%
}]{%
Ginzburg1994}%
\APACinsertmetastar{%
Ginzburg1994}%
Ginzburg, I.%
\BCBT{}\ \BBA{} Sompolinsky, H.%
%
\unskip\
\newblock
\APACrefYearMonthDay{1994}{}{}.
\newblock
\BBOQ{}\APACrefatitle{Theory of correlations in stochastic neural
  networks}{Theory of correlations in stochastic neural networks}.\BBCQ{}
\newblock
\APACjournalVolNumPages{Physical Review E}{50}{}{3171--3191}.
\PrintBackRefs{\CurrentBib}

\bibitem[\protect\citeauthoryear{%
Gollisch%
\ \BBA{} Meister%
}{%
Gollisch%
\ \BBA{} Meister%
}{%
{\protect\APACyear{2008}}%
}]{%
GollischMeister2008}%
\APACinsertmetastar{%
GollischMeister2008}%
Gollisch, T.%
\BCBT{}\ \BBA{} Meister, M.%
%
\unskip\
\newblock
\APACrefYearMonthDay{2008}{}{}.
\newblock
\BBOQ{}\APACrefatitle{Rapid neural coding in the retina with relative spike
  latencies}{Rapid neural coding in the retina with relative spike
  latencies}.\BBCQ{}
\newblock
\APACjournalVolNumPages{Science}{319}{}{1108--1111}.
\PrintBackRefs{\CurrentBib}

\bibitem[\protect\citeauthoryear{%
Hopfield%
}{%
Hopfield%
}{%
{\protect\APACyear{1995}}%
}]{%
Hopfield1995}%
\APACinsertmetastar{%
Hopfield1995}%
Hopfield, J\BPBI J.%
%
\unskip\
\newblock
\APACrefYearMonthDay{1995}{}{}.
\newblock
\BBOQ{}\APACrefatitle{Pattern recognition computation using action potential
  timing for stimulus representation}{Pattern recognition computation using
  action potential timing for stimulus representation}.\BBCQ{}
\newblock
\APACjournalVolNumPages{Nature}{376}{}{33--36}.
\PrintBackRefs{\CurrentBib}

\bibitem[\protect\citeauthoryear{%
Hosoya%
, Baccus%
\BCBL{}\ \BBA{} Meister%
}{%
Hosoya%
\ \protect\BOthers{.}}{%
{\protect\APACyear{2005}}%
}]{%
Hosoya2005}%
\APACinsertmetastar{%
Hosoya2005}%
Hosoya, T.%
, Baccus, S\BPBI A.%
\BCBL{}\ \BBA{} Meister, M.%
%
\unskip\
\newblock
\APACrefYearMonthDay{2005}{}{}.
\newblock

\newblock
\APACjournalVolNumPages{Nature}{436}{}{71--77}.
\PrintBackRefs{\CurrentBib}

\bibitem[\protect\citeauthoryear{%
Jung%
}{%
Jung%
}{%
{\protect\APACyear{1994}}%
}]{%
Jung1994}%
\APACinsertmetastar{%
Jung1994}%
Jung, P.%
%
\unskip\
\newblock
\APACrefYearMonthDay{1994}{}{}.
\newblock
\BBOQ{}\APACrefatitle{Threshold devices: Fractal noise and neural
  talk}{Threshold devices: Fractal noise and neural talk}.\BBCQ{}
\newblock
\APACjournalVolNumPages{Physical Review E}{50}{}{2513--2522}.
\PrintBackRefs{\CurrentBib}

\bibitem[\protect\citeauthoryear{%
Kanev%
, Wenning%
\BCBL{}\ \BBA{} Obermayer%
}{%
Kanev%
\ \protect\BOthers{.}}{%
{\protect\APACyear{2003}}%
}]{%
KanevWenningObermayer2004}%
\APACinsertmetastar{%
KanevWenningObermayer2004}%
Kanev, J.%
, Wenning, G.%
\BCBL{}\ \BBA{} Obermayer, K.%
%
\unskip\
\newblock
\APACrefYearMonthDay{2003}{}{}.
\newblock
\BBOQ{}\APACrefatitle{Approximating the response stimulus correlation for the
  integrate-and-fire neuron}{Approximating the response stimulus correlation
  for the integrate-and-fire neuron}.\BBCQ{}
\newblock
\APACjournalVolNumPages{Neurocomputing}{58}{}{47--52}.
\PrintBackRefs{\CurrentBib}

\bibitem[\protect\citeauthoryear{%
Keat%
, Reinagel%
, Reid%
\BCBL{}\ \BBA{} Meister%
}{%
Keat%
\ \protect\BOthers{.}}{%
{\protect\APACyear{2001}}%
}]{%
Predicting}%
\APACinsertmetastar{%
Predicting}%
Keat, J.%
, Reinagel, P.%
, Reid, R\BPBI C.%
\BCBL{}\ \BBA{} Meister, M.%
%
\unskip\
\newblock
\APACrefYearMonthDay{2001}{}{}.
\newblock
\BBOQ{}\APACrefatitle{Predicting every spike: A model for the responses of
  visual neurons}{Predicting every spike: A model for the responses of visual
  neurons}.\BBCQ{}
\newblock
\APACjournalVolNumPages{Neuron}{30}{}{803--817}.
\PrintBackRefs{\CurrentBib}

\bibitem[\protect\citeauthoryear{%
Korenberg%
\ \BBA{} Hunter%
}{%
Korenberg%
\ \BBA{} Hunter%
}{%
{\protect\APACyear{1986}}%
}]{%
Korenberg1986}%
\APACinsertmetastar{%
Korenberg1986}%
Korenberg, M\BPBI J.%
\BCBT{}\ \BBA{} Hunter, I\BPBI W.%
%
\unskip\
\newblock
\APACrefYearMonthDay{1986}{}{}.
\newblock
\BBOQ{}\APACrefatitle{The identification of nonlinear biological systems: LNL
  cascade models}{The identification of nonlinear biological systems: Lnl
  cascade models}.\BBCQ{}
\newblock
\APACjournalVolNumPages{Biological Cybernetics}{55}{}{125--134}.
\PrintBackRefs{\CurrentBib}

\bibitem[\protect\citeauthoryear{%
Kuhn%
, Aertsen%
\BCBL{}\ \BBA{} Rotter%
}{%
Kuhn%
\ \protect\BOthers{.}}{%
{\protect\APACyear{2003}}%
}]{%
KuhnAertsenRotter2003}%
\APACinsertmetastar{%
KuhnAertsenRotter2003}%
Kuhn, A.%
, Aertsen, A.%
\BCBL{}\ \BBA{} Rotter, S.%
%
\unskip\
\newblock
\APACrefYearMonthDay{2003}{}{}.
\newblock
\BBOQ{}\APACrefatitle{Higher-order statistics of input ensembles and the
  response of simple model neurons}{Higher-order statistics of input ensembles
  and the response of simple model neurons}.\BBCQ{}
\newblock
\APACjournalVolNumPages{Neural Computation}{15}{}{67--101}.
\PrintBackRefs{\CurrentBib}

\bibitem[\protect\citeauthoryear{%
L.%
, Gerstner%
\BCBL{}\ \BBA{} Richardson%
}{%
L.%
\ \protect\BOthers{.}}{%
{\protect\APACyear{2006}}%
}]{%
Badel2006}%
\APACinsertmetastar{%
Badel2006}%
L., B.%
, Gerstner, W.%
\BCBL{}\ \BBA{} Richardson, M\BPBI J.%
%
\unskip\
\newblock
\APACrefYearMonthDay{2006}{}{}.
\newblock
\BBOQ{}\APACrefatitle{Dependence of the spike-triggered average voltage on
  membrane response properties}{Dependence of the spike-triggered average
  voltage on membrane response properties}.\BBCQ{}
\newblock
\APACjournalVolNumPages{Neurocomputing}{69}{}{1062--1065}.
\PrintBackRefs{\CurrentBib}

\bibitem[\protect\citeauthoryear{%
Mainen%
\ \BBA{} Sejnowski%
}{%
Mainen%
\ \BBA{} Sejnowski%
}{%
{\protect\APACyear{1995}}%
}]{%
MainenSejnowski1995}%
\APACinsertmetastar{%
MainenSejnowski1995}%
Mainen, Z\BPBI F.%
\BCBT{}\ \BBA{} Sejnowski, T\BPBI J.%
%
\unskip\
\newblock
\APACrefYearMonthDay{1995}{}{}.
\newblock
\BBOQ{}\APACrefatitle{Reliability of spike timing in neocortical
  neurons}{Reliability of spike timing in neocortical neurons}.\BBCQ{}
\newblock
\APACjournalVolNumPages{Science}{268}{}{1503--1506}.
\PrintBackRefs{\CurrentBib}

\bibitem[\protect\citeauthoryear{%
Markowitz%
, Collman%
, Brody%
\BCBL{}\ \BBA{} Hopfield%
}{%
Markowitz%
\ \protect\BOthers{.}}{%
{\protect\APACyear{2008}}%
}]{%
Markowitz2008}%
\APACinsertmetastar{%
Markowitz2008}%
Markowitz, D\BPBI A.%
, Collman, F.%
, Brody, C\BPBI D.%
\BCBL{}\ \BBA{} Hopfield, D\BPBI W., J. J.~andTank.%
%
\unskip\
\newblock
\APACrefYearMonthDay{2008}{}{}.
\newblock
\BBOQ{}\APACrefatitle{Rate-specific synchrony: Using noisy oscillations to
  detect equally active neurons}{Rate-specific synchrony: Using noisy
  oscillations to detect equally active neurons}.\BBCQ{}
\newblock
\APACjournalVolNumPages{Proceedings of the National Academy of Sciences of the
  USA}{105}{}{8422--8427}.
\PrintBackRefs{\CurrentBib}

\bibitem[\protect\citeauthoryear{%
Meister%
\ \BBA{} Berry%
}{%
Meister%
\ \BBA{} Berry%
}{%
{\protect\APACyear{1999}}%
}]{%
MeisterBerry1999}%
\APACinsertmetastar{%
MeisterBerry1999}%
Meister, M.%
\BCBT{}\ \BBA{} Berry, M\BPBI J.%
%
\unskip\
\newblock
\APACrefYearMonthDay{1999}{}{}.
\newblock
\BBOQ{}\APACrefatitle{The neural code of the retina}{The neural code of the
  retina}.\BBCQ{}
\newblock
\APACjournalVolNumPages{Neuron}{22}{}{435--450}.
\PrintBackRefs{\CurrentBib}

\bibitem[\protect\citeauthoryear{%
Moreno%
, Rocha%
, Renart%
\BCBL{}\ \BBA{} Parga%
}{%
Moreno%
\ \protect\BOthers{.}}{%
{\protect\APACyear{2002}}%
}]{%
MorenoRochaRenartParga2002}%
\APACinsertmetastar{%
MorenoRochaRenartParga2002}%
Moreno, R.%
, Rocha, J. de~la%
, Renart, A.%
\BCBL{}\ \BBA{} Parga, N.%
%
\unskip\
\newblock
\APACrefYearMonthDay{2002}{}{}.
\newblock
\BBOQ{}\APACrefatitle{Response of spiking neurons to correlated
  inputs}{Response of spiking neurons to correlated inputs}.\BBCQ{}
\newblock
\APACjournalVolNumPages{Physical Review Letters}{89}{}{288101}.
\PrintBackRefs{\CurrentBib}

\bibitem[\protect\citeauthoryear{%
Moreno-Bote%
\ \BBA{} Parga%
}{%
Moreno-Bote%
\ \BBA{} Parga%
}{%
{\protect\APACyear{2006}}%
}]{%
MorenoParga2006}%
\APACinsertmetastar{%
MorenoParga2006}%
Moreno-Bote, R.%
\BCBT{}\ \BBA{} Parga, N.%
%
\unskip\
\newblock
\APACrefYearMonthDay{2006}{}{}.
\newblock
\BBOQ{}\APACrefatitle{Auto- and crosscorrelograms for the spike response of
  leaky integrate-and-fire neurons with slow synapses}{Auto- and
  crosscorrelograms for the spike response of leaky integrate-and-fire neurons
  with slow synapses}.\BBCQ{}
\newblock
\APACjournalVolNumPages{Physical Review Letters}{96}{}{028101}.
\PrintBackRefs{\CurrentBib}

\bibitem[\protect\citeauthoryear{%
Paninski%
}{%
Paninski%
}{%
{\protect\APACyear{2003}}%
}]{%
Paninski2003}%
\APACinsertmetastar{%
Paninski2003}%
Paninski, L.%
%
\unskip\
\newblock
\APACrefYearMonthDay{2003}{}{}.
\newblock
\BBOQ{}\APACrefatitle{Convergence properties of three spike-triggered analysis
  techniques}{Convergence properties of three spike-triggered analysis
  techniques}.\BBCQ{}
\newblock
\APACjournalVolNumPages{Network: Computing in Neural Systems}{14}{}{437--464}.
\PrintBackRefs{\CurrentBib}

\bibitem[\protect\citeauthoryear{%
Paninski%
}{%
Paninski%
}{%
{\protect\APACyear{2006}}%
}]{%
Paninski2006}%
\APACinsertmetastar{%
Paninski2006}%
Paninski, L.%
%
\unskip\
\newblock
\APACrefYearMonthDay{2006}{}{}.
\newblock
\BBOQ{}\APACrefatitle{The spike-triggered average of the integrate-and-fire
  cell driven by Gaussian white noise}{The spike-triggered average of the
  integrate-and-fire cell driven by gaussian white noise}.\BBCQ{}
\newblock
\APACjournalVolNumPages{Neural Computation}{18}{}{2592--2616}.
\PrintBackRefs{\CurrentBib}

\bibitem[\protect\citeauthoryear{%
Rice%
}{%
Rice%
}{%
{\protect\APACyear{1954}}%
}]{%
Rice}%
\APACinsertmetastar{%
Rice}%
Rice, S\BPBI O.%
%
\unskip\
\newblock
\APACrefYearMonthDay{1954}{}{}.
\newblock
\BBOQ{}\APACrefatitle{Mathematical Analysis of Random Noise}{Mathematical
  analysis of random noise}.\BBCQ{}
\newblock
\BIn{} N.~Wax\ (\BED), \APACrefbtitle{Selected Papers on Noise and Stochastic
  Processes}{Selected papers on noise and stochastic processes}\ (\BPGS\
  133--294).
\newblock
\APACaddressPublisher{New York}{Dover}.
\PrintBackRefs{\CurrentBib}

\bibitem[\protect\citeauthoryear{%
Rieke%
}{%
Rieke%
}{%
{\protect\APACyear{2001}}%
}]{%
Rieke2001}%
\APACinsertmetastar{%
Rieke2001}%
Rieke, F.%
%
\unskip\
\newblock
\APACrefYearMonthDay{2001}{}{}.
\newblock
\BBOQ{}\APACrefatitle{Temporal contrast adaptation in salamander bipolar
  cells}{Temporal contrast adaptation in salamander bipolar cells}.\BBCQ{}
\newblock
\APACjournalVolNumPages{Journal of Neuroscience}{21}{}{9445--9454}.
\PrintBackRefs{\CurrentBib}

\bibitem[\protect\citeauthoryear{%
Rieke%
, Warland%
, Stevennick%
\BCBL{}\ \BBA{} Bialek%
}{%
Rieke%
\ \protect\BOthers{.}}{%
{\protect\APACyear{1996}}%
}]{%
Spikes}%
\APACinsertmetastar{%
Spikes}%
Rieke, F.%
, Warland, D.%
, Stevennick, R. van%
\BCBL{}\ \BBA{} Bialek, W.%
%
\unskip\
\newblock
\APACrefYear{1996}.
\newblock
\APACrefbtitle{Spikes: Exploring the Neural Code}{Spikes: Exploring the neural
  code}.
\newblock
\APACaddressPublisher{Cambridge, USA}{MIT Press}.
\PrintBackRefs{\CurrentBib}

\bibitem[\protect\citeauthoryear{%
Rust%
, Scwartz%
, Movshon%
\BCBL{}\ \BBA{} Simoncelli%
}{%
Rust%
\ \protect\BOthers{.}}{%
{\protect\APACyear{2005}}%
}]{%
Rust2005}%
\APACinsertmetastar{%
Rust2005}%
Rust, N\BPBI C.%
, Scwartz, O.%
, Movshon, J\BPBI A.%
\BCBL{}\ \BBA{} Simoncelli, E\BPBI P.%
%
\unskip\
\newblock
\APACrefYearMonthDay{2005}{}{}.
\newblock
\BBOQ{}\APACrefatitle{Spatiotemporal elements of macaque V1 receptive
  fields}{Spatiotemporal elements of macaque v1 receptive fields}.\BBCQ{}
\newblock
\APACjournalVolNumPages{Neuron}{46}{}{945--956}.
\PrintBackRefs{\CurrentBib}

\bibitem[\protect\citeauthoryear{%
Salinas%
\ \BBA{} Sejnowski%
}{%
Salinas%
\ \BBA{} Sejnowski%
}{%
{\protect\APACyear{2002}}%
}]{%
SalinasSejnowski2002}%
\APACinsertmetastar{%
SalinasSejnowski2002}%
Salinas, E.%
\BCBT{}\ \BBA{} Sejnowski, T.%
%
\unskip\
\newblock
\APACrefYearMonthDay{2002}{}{}.
\newblock
\BBOQ{}\APACrefatitle{Integrate-and-fire neurons driven by correlated
  stochastic input}{Integrate-and-fire neurons driven by correlated stochastic
  input}.\BBCQ{}
\newblock
\APACjournalVolNumPages{Neural Computation}{14}{}{2111--2155}.
\PrintBackRefs{\CurrentBib}

\bibitem[\protect\citeauthoryear{%
Salinas%
\ \BBA{} Sejnowski%
}{%
Salinas%
\ \BBA{} Sejnowski%
}{%
{\protect\APACyear{2000}}%
}]{%
SalinasSejnowski2000}%
\APACinsertmetastar{%
SalinasSejnowski2000}%
Salinas, E.%
\BCBT{}\ \BBA{} Sejnowski, T\BPBI J.%
%
\unskip\
\newblock
\APACrefYearMonthDay{2000}{}{}.
\newblock
\BBOQ{}\APACrefatitle{Impact of correlated synaptic input on output firing rate
  and variability in simple neuronal models}{Impact of correlated synaptic
  input on output firing rate and variability in simple neuronal
  models}.\BBCQ{}
\newblock
\APACjournalVolNumPages{Journal of Neuroscience}{20}{}{6193--6209}.
\PrintBackRefs{\CurrentBib}

\bibitem[\protect\citeauthoryear{%
Schwalger%
\ \BBA{} Schimansky-Geier%
}{%
Schwalger%
\ \BBA{} Schimansky-Geier%
}{%
{\protect\APACyear{2008}}%
}]{%
SchwalgerSchimanskyGeier2008}%
\APACinsertmetastar{%
SchwalgerSchimanskyGeier2008}%
Schwalger, T.%
\BCBT{}\ \BBA{} Schimansky-Geier, L.%
%
\unskip\
\newblock
\APACrefYearMonthDay{2008}{}{}.
\newblock
\BBOQ{}\APACrefatitle{Interspike interval statistics of a leaky
  integrate-and-fire neuron driven by Gaussian noise with large correlation
  times}{Interspike interval statistics of a leaky integrate-and-fire neuron
  driven by gaussian noise with large correlation times}.\BBCQ{}
\newblock
\APACjournalVolNumPages{Physical Review E}{77}{}{031914}.
\PrintBackRefs{\CurrentBib}

\bibitem[\protect\citeauthoryear{%
Schwartz%
}{%
Schwartz%
}{%
{\protect\APACyear{2006}}%
}]{%
Schwartz2006}%
\APACinsertmetastar{%
Schwartz2006}%
Schwartz, O.%
%
\unskip\
\newblock
\APACrefYearMonthDay{2006}{}{}.
\newblock
\BBOQ{}\APACrefatitle{Spike-triggered neural characterization}{Spike-triggered
  neural characterization}.\BBCQ{}
\newblock
\APACjournalVolNumPages{Journal of Vision}{6}{}{484--507}.
\PrintBackRefs{\CurrentBib}

\bibitem[\protect\citeauthoryear{%
R.~de~Ruyter~van Steveninck%
\ \BBA{} Bialek%
}{%
R.~de~Ruyter~van Steveninck%
\ \BBA{} Bialek%
}{%
{\protect\APACyear{1988}}%
}]{%
BialekVanSteveninck1988}%
\APACinsertmetastar{%
BialekVanSteveninck1988}%
Steveninck, R. de~Ruyter~van%
\BCBT{}\ \BBA{} Bialek, W.%
%
\unskip\
\newblock
\APACrefYearMonthDay{1988}{}{}.
\newblock
\BBOQ{}\APACrefatitle{Coding and information transfer in short spike
  sequences}{Coding and information transfer in short spike sequences}.\BBCQ{}
\newblock
\APACjournalVolNumPages{Proceedings of the Royal Society of London
  B}{234}{}{379--414}.
\PrintBackRefs{\CurrentBib}

\bibitem[\protect\citeauthoryear{%
R\BPBI R.~de~Ruyter~van Steveninck%
, Lewen%
, Strong%
, Koberle%
\BCBL{}\ \BBA{} Bialek%
}{%
R\BPBI R.~de~Ruyter~van Steveninck%
\ \protect\BOthers{.}}{%
{\protect\APACyear{1997}}%
}]{%
Bialek1997}%
\APACinsertmetastar{%
Bialek1997}%
Steveninck, R\BPBI R. de~Ruyter~van%
, Lewen, G\BPBI D.%
, Strong, S\BPBI P.%
, Koberle, R.%
\BCBL{}\ \BBA{} Bialek, W.%
%
\unskip\
\newblock
\APACrefYearMonthDay{1997}{}{}.
\newblock
\BBOQ{}\APACrefatitle{Reproducibility and variability in neural spike
  trains}{Reproducibility and variability in neural spike trains}.\BBCQ{}
\newblock
\APACjournalVolNumPages{Science}{275}{}{1805--1808}.
\PrintBackRefs{\CurrentBib}

\bibitem[\protect\citeauthoryear{%
Stroeve%
\ \BBA{} Gielen%
}{%
Stroeve%
\ \BBA{} Gielen%
}{%
{\protect\APACyear{2001}}%
}]{%
StroeveGielen2001}%
\APACinsertmetastar{%
StroeveGielen2001}%
Stroeve, S.%
\BCBT{}\ \BBA{} Gielen, S.%
%
\unskip\
\newblock
\APACrefYearMonthDay{2001}{}{}.
\newblock
\BBOQ{}\APACrefatitle{Correlation between uncoupled conductance based
  integrate-and-fire neurons due to common and syncrhonous presynaptic
  firing}{Correlation between uncoupled conductance based integrate-and-fire
  neurons due to common and syncrhonous presynaptic firing}.\BBCQ{}
\newblock
\APACjournalVolNumPages{Neural Computation}{13}{}{2005--2029}.
\PrintBackRefs{\CurrentBib}

\bibitem[\protect\citeauthoryear{%
Tchumatchenko%
, Malyshev%
, Geisel%
, Volgushev%
\BCBL{}\ \BBA{} Wolf%
}{%
Tchumatchenko%
\ \protect\BOthers{.}}{%
{\protect\APACyear{2008}}%
}]{%
Tchumatchenko2008}%
\APACinsertmetastar{%
Tchumatchenko2008}%
Tchumatchenko, T.%
, Malyshev, A.%
, Geisel, T.%
, Volgushev, M.%
\BCBL{}\ \BBA{} Wolf, F.%
%
\unskip\
\newblock
\APACrefYearMonthDay{2008}{}{}.
\newblock
\BBOQ{}\APACrefatitle{Correlations and synchrony in threshold neuron
  models}{Correlations and synchrony in threshold neuron models}.\BBCQ{}
\newblock
\APAChowpublished{Available online at http://arxiv.org/abs/0810.2901v2.}
\PrintBackRefs{\CurrentBib}

\bibitem[\protect\citeauthoryear{%
Tuckwell%
}{%
Tuckwell%
}{%
{\protect\APACyear{1988}}%
{\protect\APACexlab{{\protect\BCnt{1}}}}}]{%
Tuckwell1988a}%
\APACinsertmetastar{%
Tuckwell1988a}%
Tuckwell, H.%
%
\unskip\
\newblock
\APACrefYear{1988{\protect\BCnt{1}}}.
\newblock
\APACrefbtitle{Introduction to theoretical neurobiology: Vol. 1, Linear cable
  theory and dendritic structure}{Introduction to theoretical neurobiology:
  Vol. 1, linear cable theory and dendritic structure}.
\newblock
\APACaddressPublisher{Cambridge, UK}{Cambridge University Press}.
\PrintBackRefs{\CurrentBib}

\bibitem[\protect\citeauthoryear{%
Tuckwell%
}{%
Tuckwell%
}{%
{\protect\APACyear{1988}}%
{\protect\APACexlab{{\protect\BCnt{2}}}}}]{%
Tuckwell1988b}%
\APACinsertmetastar{%
Tuckwell1988b}%
Tuckwell, H.%
%
\unskip\
\newblock
\APACrefYear{1988{\protect\BCnt{2}}}.
\newblock
\APACrefbtitle{Introduction to theoretical neurobiology: Vol. 2, Nonlinear and
  stochastic theories}{Introduction to theoretical neurobiology: Vol. 2,
  nonlinear and stochastic theories}.
\newblock
\APACaddressPublisher{Cambridge, UK}{Cambridge University Press}.
\PrintBackRefs{\CurrentBib}

\bibitem[\protect\citeauthoryear{%
Uzzell%
\ \BBA{} Chichilnisky%
}{%
Uzzell%
\ \BBA{} Chichilnisky%
}{%
{\protect\APACyear{2004}}%
}]{%
UzzellChichilnisky2003}%
\APACinsertmetastar{%
UzzellChichilnisky2003}%
Uzzell, V.%
\BCBT{}\ \BBA{} Chichilnisky, E.%
%
\unskip\
\newblock
\APACrefYearMonthDay{2004}{}{}.
\newblock
\BBOQ{}\APACrefatitle{Precision of spike trains in primate retinal ganglion
  cells}{Precision of spike trains in primate retinal ganglion cells}.\BBCQ{}
\newblock
\APACjournalVolNumPages{Journal of Neurophysiology}{92}{}{780--789}.
\PrintBackRefs{\CurrentBib}

\end{thebibliography}
%

%
\pagebreak

\begin{figure}[tb]
\centerline{
\includegraphics[scale=1.0]{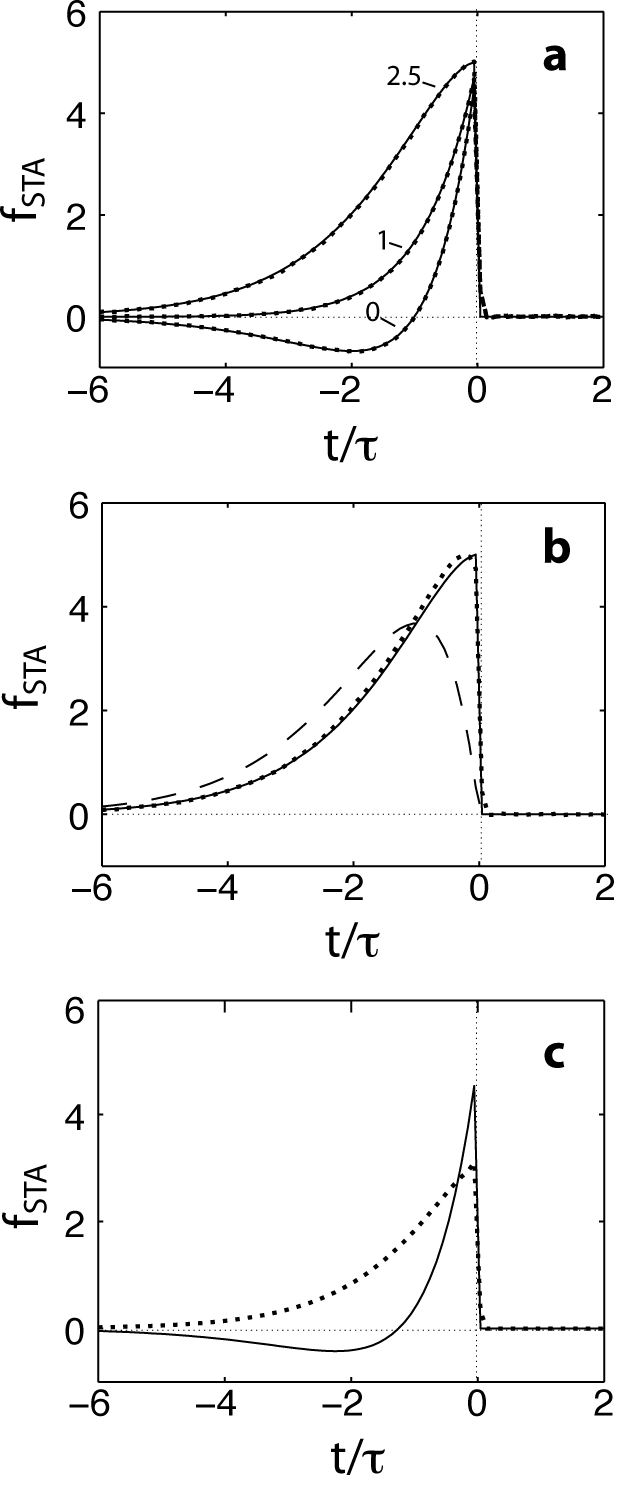}}
\caption{(a) STA
of a model neuron spiking in response to upward threshold crossing of the generating potential 
described by Eqs.~(\ref{eq:g_f}), 
(\ref{eq:falpha}), and (\ref{eq:walpha}). 
The STA, Eq.~(\ref{eq:sta_eval}), 
is plotted for three values of the threshold: $\theta = 2.5 \sigma$, $\sigma$, and $0$,
where $\sigma$ is the standard deviation of $g(t)$ (solid line: analytic expression, dotted lines: simulation).
(b) A comparison of the STA for $\theta = 2.5 \sigma$ (solid line) with simulation results
from a leaky integrate-and-fire neuron
with time constants $\tau_1 = \tau_2 = \tau$ and the same threshold, receiving $s$ as its input, and resetting its 
membrane potential to $0$ after each spike (dotted line). The dashed line shows the first term only of
Eq.~(\ref{eq:sta_eval}). (c) A similar comparison as in (b), with $\theta = 0.25 \sigma$. (solid line, threshold crossing
model, dashed line, leaky integrate-and-fire model.)}
\end{figure}
\pagebreak
\begin{figure*}[tb]
\centerline{
\includegraphics[scale=1.0]{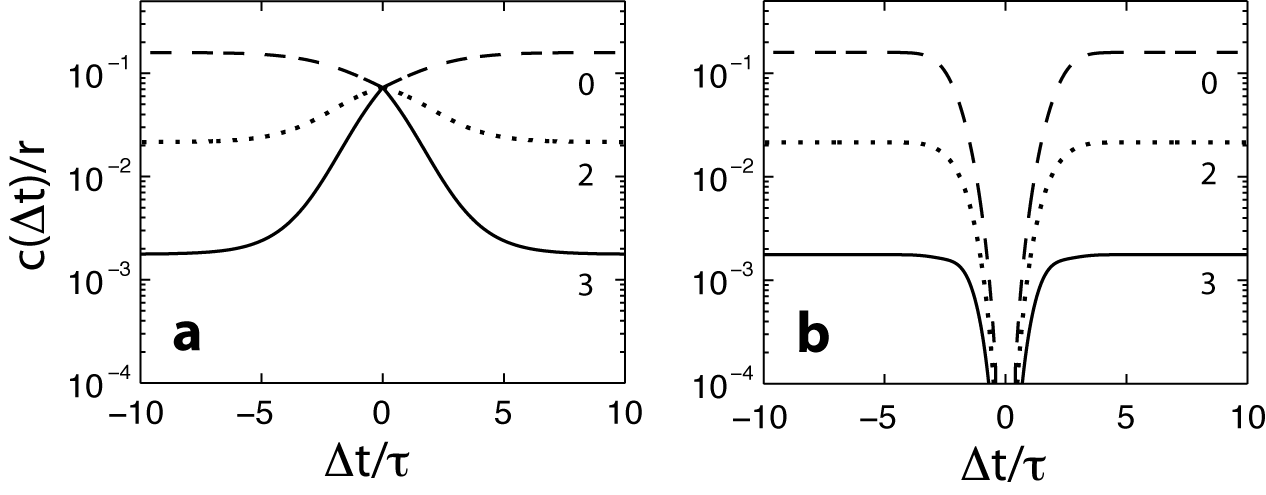}}
\caption{Spike auto-correlation function, $c(\Delta t)$, divided by the average firing rate $r$,
for three values of the threshold: $\theta = 0$, $2\sigma$, and $3\sigma$. The
generating potential is described by
Eqs.~(\ref{eq:g_f}), 
(\ref{eq:falpha}), and (\ref{eq:walpha}). (b) A similar plot for a generating potential having
a Gaussian correlation function, $w(t) = \sigma^2{\rm exp}(-t^2/2 \tau^2)$. }
\vspace*{2in}
\end{figure*}
\begin{figure*}[tb]
\centerline{
\includegraphics[scale=0.8]{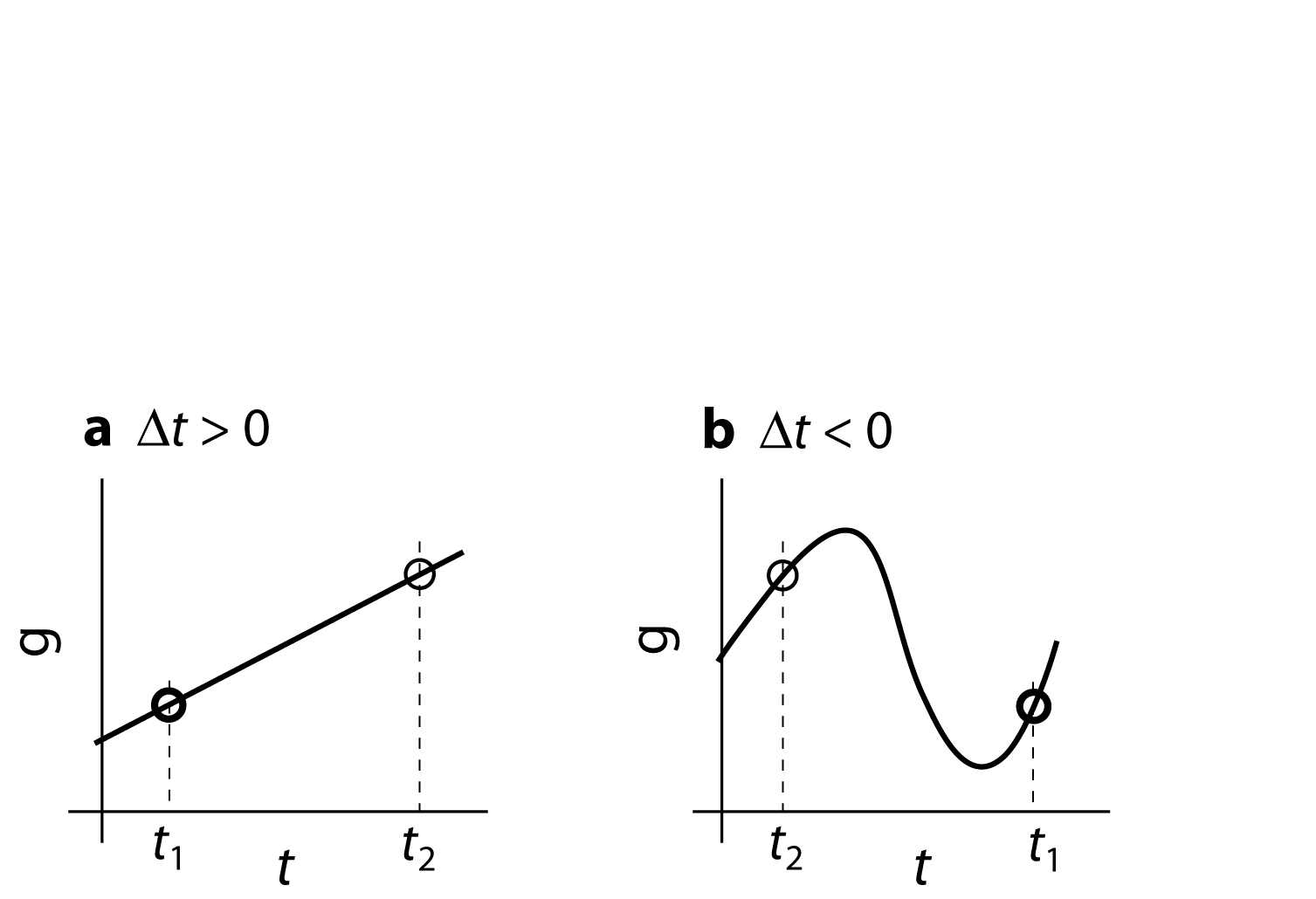}}
\caption{Spike generation events in the threshold crossing model, for two neurons with
thresholds $\theta_1$ (thick circle) and $\theta_2 > \theta_1$ (thin circle). In the left plot
neuron 2 fires after neuron 1, whereas in the right plot neuron1 fires after neuron 2.
In the latter case the generating potential must reverse the sign of its
derivative twice within the time interval separating the two spikes.}
\vspace*{2in}
\end{figure*}
\pagebreak
\begin{figure*}[tb]
\centerline{
\includegraphics[scale=1.0]{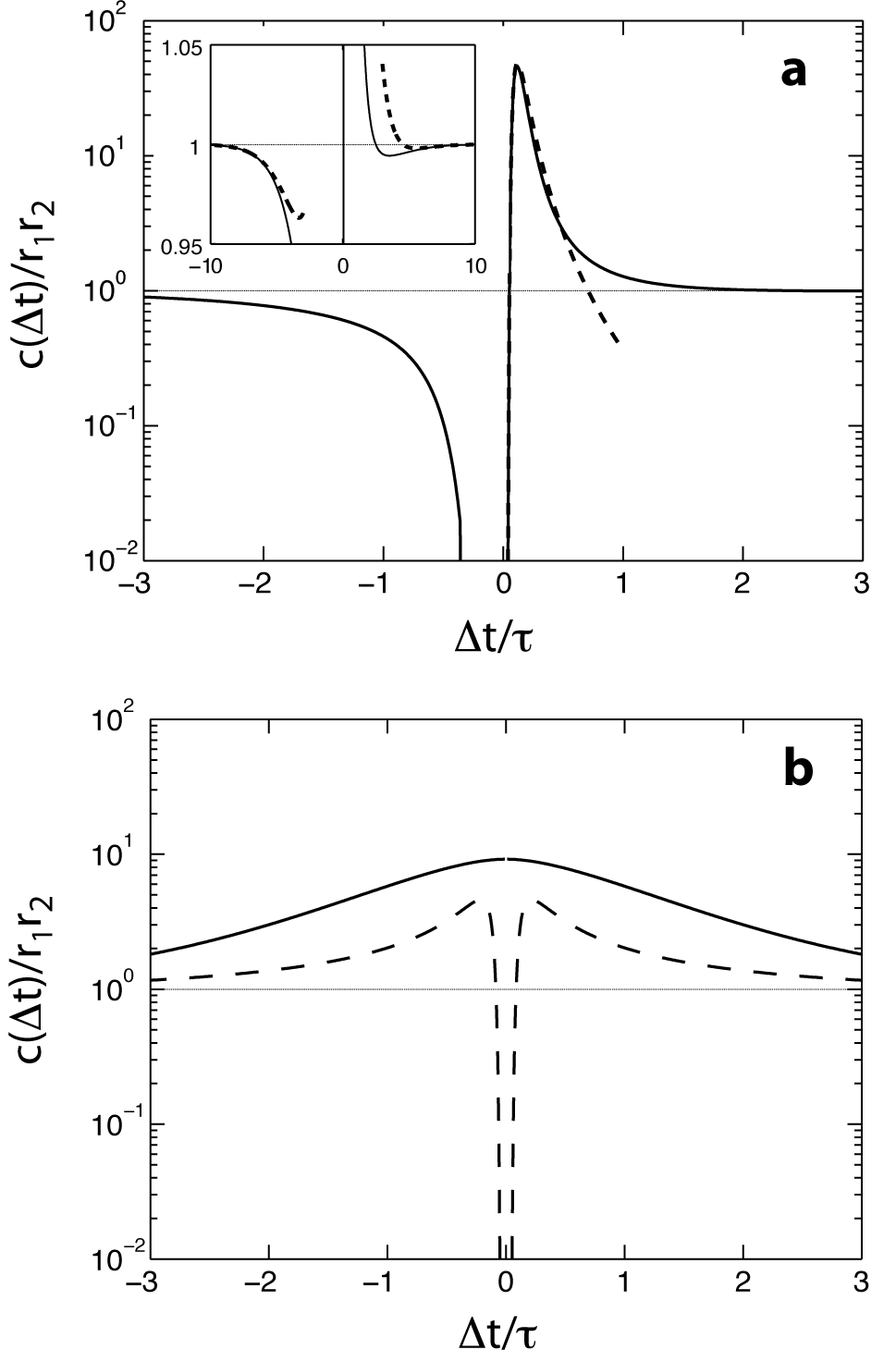}}
\caption{(a) The spike correlation function, $c(\Delta t)/r_1 r_2$ (solid line,) of two model neurons with thresholds
$\theta_1 = 0.8\sigma$ and $\theta_2 = \sigma$, firing in response to the same stimulus (same as in Fig.~2a). The dashed line 
shows the approximation for small $\Delta t$, Eqs.~(\ref{eq:I_small_t_pos_ap}) and (\ref{eq:p_tt_small_t})--(\ref{eq:gamma1}). 
At large $|\Delta t|$
the two neurons become decorrelated, and $c(\Delta t)/r_1 r_2$ approaches unity. The inset shows
the approximation for weakly correlated neurons, Eq.~(\ref{eq:spike_corr_large_t}) (dashed line) compared with the actual
correlation function (solid line). (b) The prediction of a LN model for the spike correlation
function, $c(\Delta t)/r_1 r_2$, shown for two different forms of the non-linearity, as described in the text
(linear rectification, solid line; delta function, dashed line).
}
\vspace*{2in}
\end{figure*}
\pagebreak
\begin{figure}[tb]
\centerline{
\includegraphics[scale=1.0]{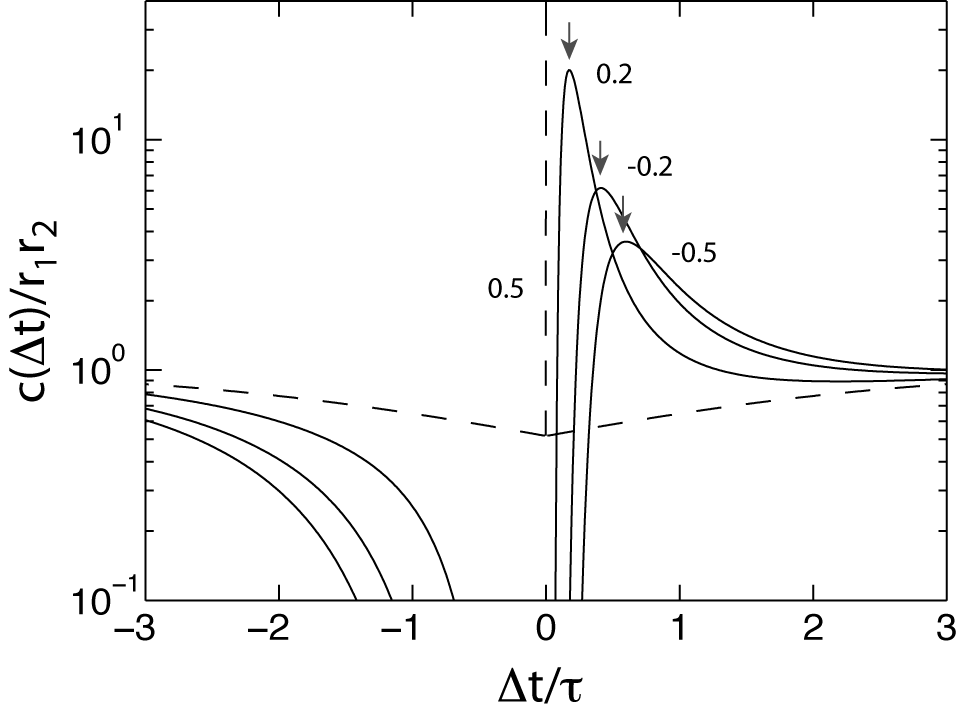}}
\caption{Spike correlation function, $c(\Delta t)/r_1 r_2$, for two model neurons with $\theta_2 = 0.5\sigma$ and
$\theta_1 = 0.5\sigma$ (dashed line), $0.2\sigma$, $-0.2\sigma$, and $-0.5\sigma$ (solid lines). 
The generating potential is described by
Eqs.~(\ref{eq:g_f}), 
(\ref{eq:f}), and (\ref{eq:f}). Arrows point to the position of the peak according
to the prediction of Eq.~(\ref{eq:latency}).
}
\vspace*{2in}
\end{figure}
\pagebreak
\begin{figure*}[tb]
\centerline{
\includegraphics[scale=1.0]{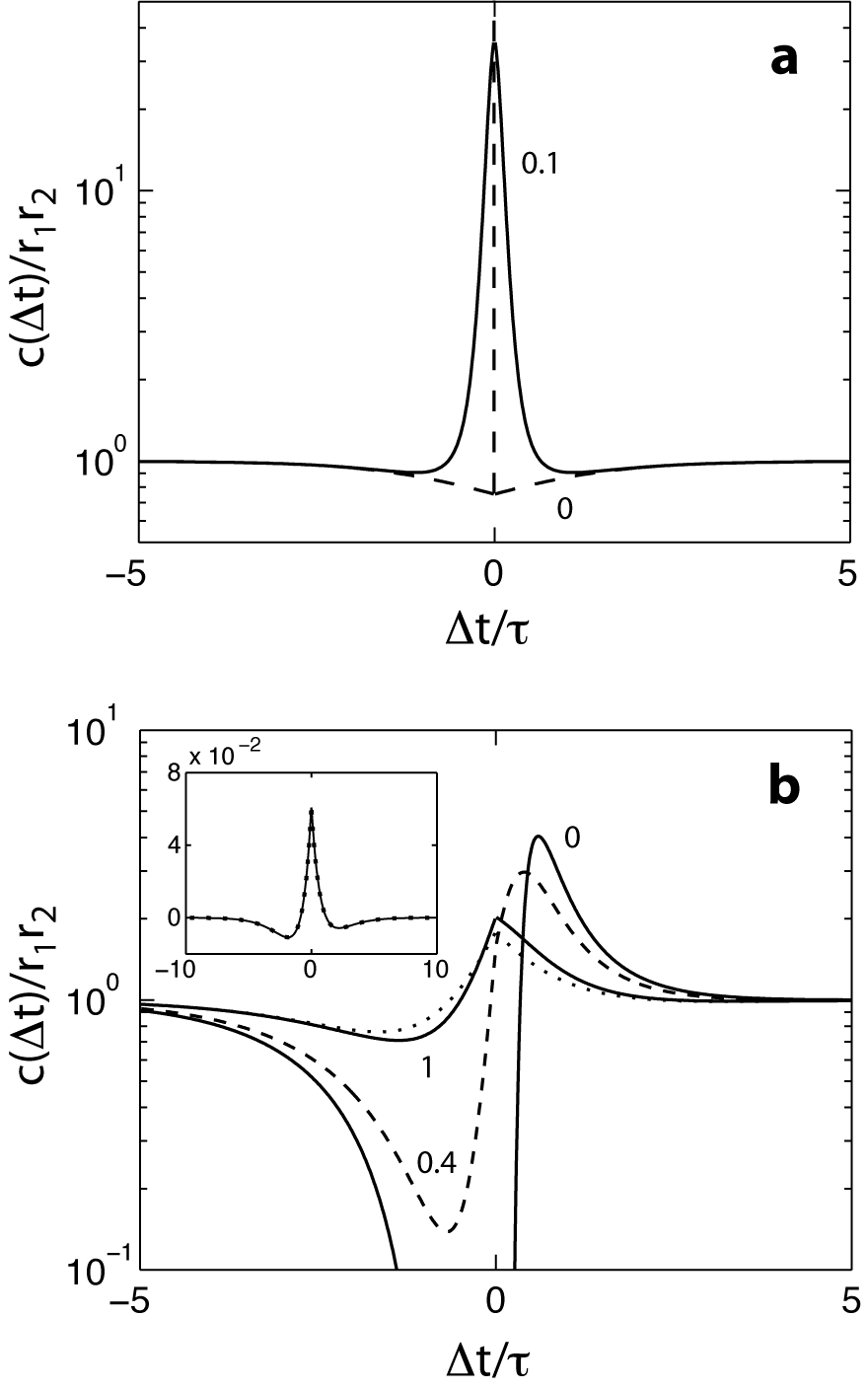}}
\caption{(a) Spike correlation function, $c(\Delta t)/r_1 r_2$,
of two model neurons with identical thresholds
$\theta = \sigma$, receiving an identical stimulus and uncorrelated noise,
Eq.~(\ref{eq:g_f_s_xi})
(solid line.)
The standard deviation of the noise, divided by that of the common stimulus, is $\alpha = 0.1$ 
(other parameters are described in the text.)
This is compared with the case where there is no noise, $\alpha = 0$ (dashed line.) (b) Spike correlation function 
of two neurons receiving similar input as in (a), but differing in threshold: $\theta_1 = 0$ and $\theta_2 = \sigma$, 
where $\sigma$ is the standard deviation of the common stimulus $s$. The correlation function is plotted for
three different ratios between the noise standard deviation and the stimulus's 
standard deviation: $\alpha = 0$, $0.4$, and $1$. The inset
shows $c(\Delta t)/r_1 r_2 - 1$ in a case where the common stimulus is weak compared to the noise, $\alpha = 10$ (solid line).
The dotted line in the inset and in the main plot (for $\alpha = 1$)
is the approximation for the case of weak correlation, Eq.~(\ref{eq:weak_correlation}). }
\end{figure*}
\pagebreak
\begin{figure}[tb]
\centerline{
\includegraphics[scale=0.75]{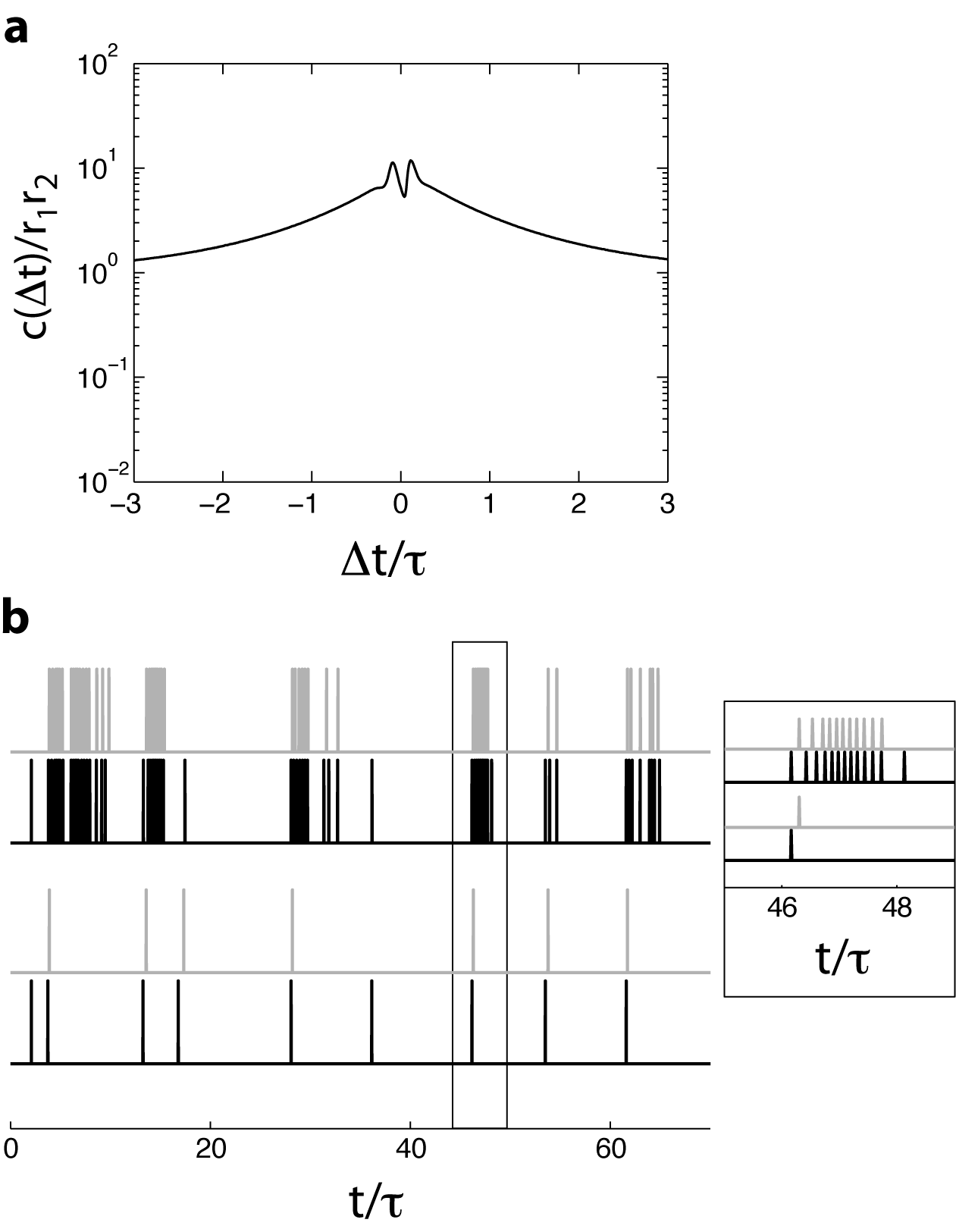}}
\caption{(a) Spike cross-correlation function of two simulated
LIF neurons, Eqs.~(\ref{eq:IAF1})--(\ref{eq:IAF2})
with $\tau_1 = \tau_2 \equiv \tau$, and
with thresholds $0.8\,\sigma$ and $1.0\,\sigma$, receiving
an identical stimulus (as described in the text, up to a baseline shift).
In both neurons $\theta - u_r = 0.5 \sigma$.
(b) Spike trains generated by the two LIF neurons (top two traces: black, $\theta_1 = 0.8$ and
gray, $\theta_2 = 1$) and by threshold crossing model neurons responding to the same 
stimulus (bottom two traces). Inset: blow-up of an interval showing a burst generated by the two LIF neurons.
}
\end{figure}
\pagebreak
\begin{figure}[tb]
\centerline{
\includegraphics[scale=0.75]{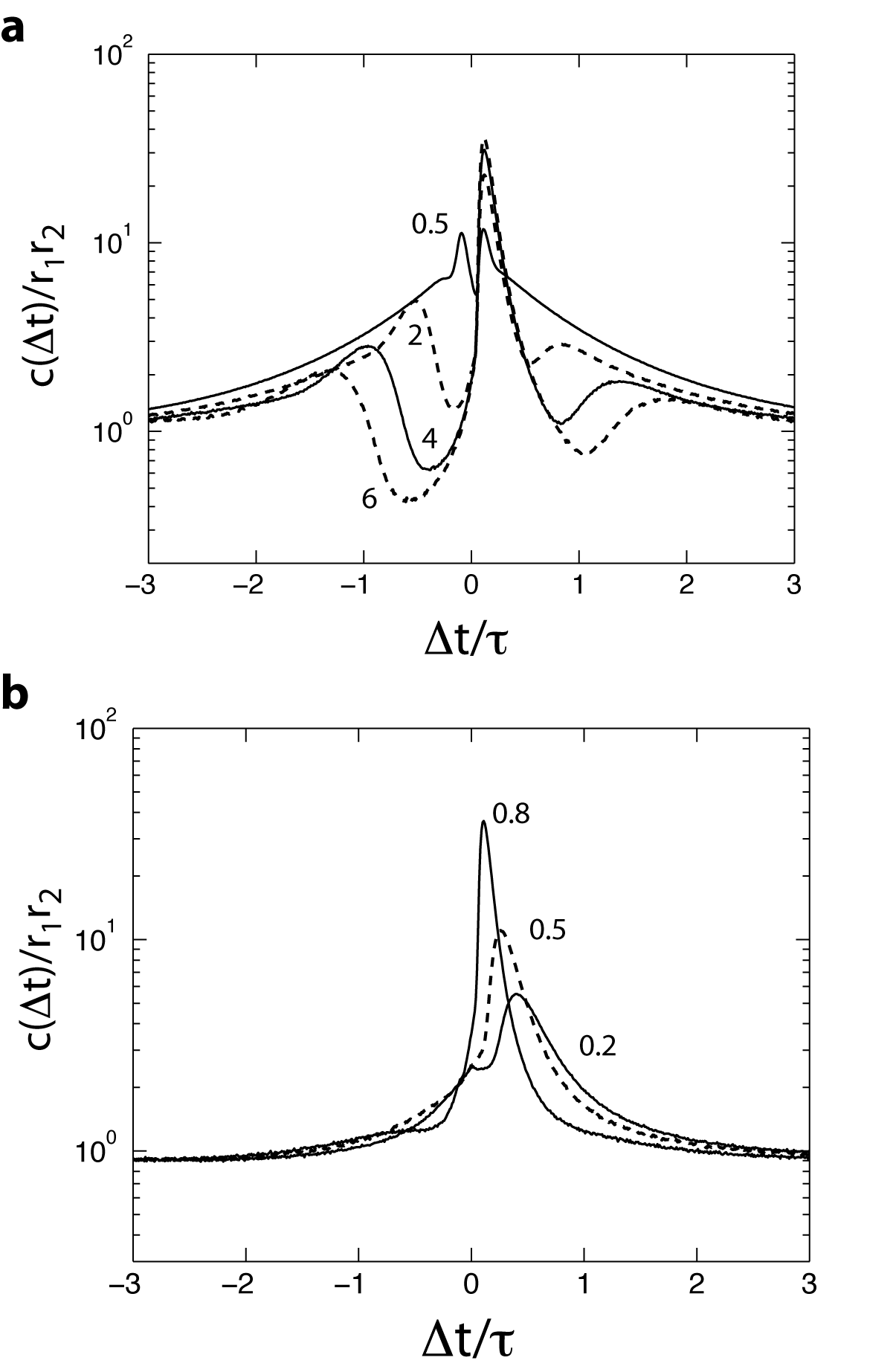}}
\caption{(a) Spike cross correlation functions of LIF neurons with 
increasing values $\theta - u_r$, the hyperpolarizing jump in potential following each spike 
(alternating solid and dashed lines:
$0.5\sigma$, $2\sigma$, $4\sigma$, and $6\sigma$. In all traces the thresholds are 
$\theta_1 = 0.8$ and gray, $\theta_2 = 1$. The trace for $\theta - u_r = 0.5\sigma$
is identical to the one in Fig.~7 (a).
(b) Spike cross-correlation functions of pairs of LIF neurons with refractoriness (as described
in the main text.) In all pairs the second neuron has a threshold $\theta_2 = 1.0$, whereas
the first neuron has a threshold $\theta_1 = $ 0.2 (solid line), 0.5 (dashed line),
and 0.8 (solid line), and $\theta-u_r = 0.5 \sigma$.
}
\end{figure}
\pagebreak
\begin{figure}[tb]
\centerline{
\includegraphics[scale=1.0]{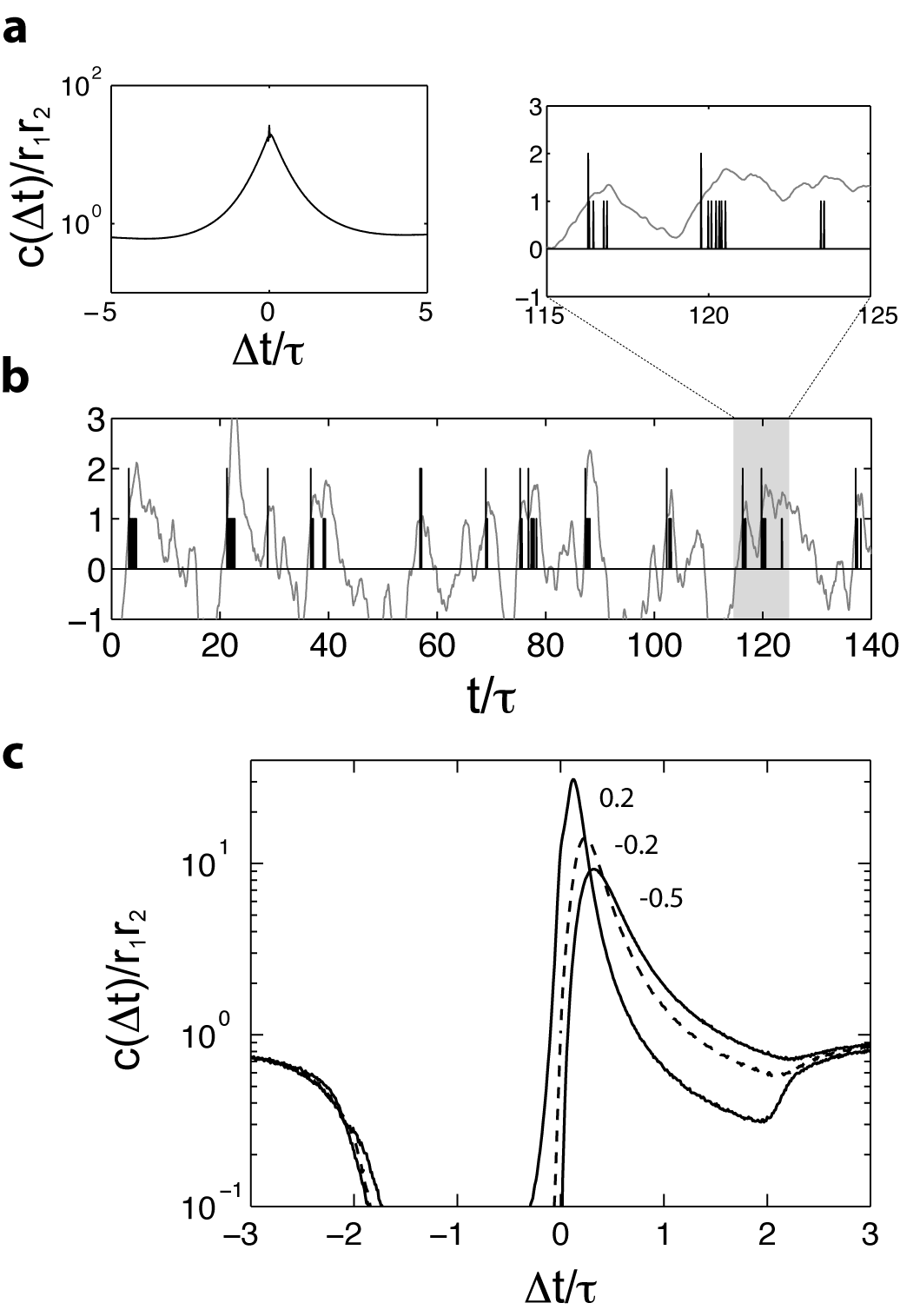}}
\caption{(a) Spike cross-correlation in the model described in Sec.~VI. The generating potential is described by Eqs.~(\ref{eq:g_f}), (\ref{eq:f}), and (\ref{eq:w}). The thresholds are $\theta_1 = 0.8 \sigma$ and $\theta_2 = \sigma$, $\tau_p = 5\tau$ and $B = 0.1$. (b) An example of a spike train generated by this model (short vertical lines). The threshold $\theta = \sigma$. For comparison, the tall vertical lines represent spike times in the threshold-crossing model. The generating potential, common to both models, is plotted in gray. Top right: blow-up of a shorter interval, showing three spiking events. (c) Cross correlation of event times in the same model, between a model neuron with threshold $\theta_1 = 0.5$ and three model neurons with thresholds $\theta_2 = 0.2$, $-0.2$, and $-0.5$. Other parameters are as in panels a and b. Event onsets are isolated by discarding any spike that occurs within a time frame of $2\tau$ from a previous spike.}
\end{figure}
\pagebreak
\end{document}